\documentclass[
    journal,
    twocolumn,
    ]{IEEEtran}
\usepackage{graphicx}
\usepackage{amsmath}
\usepackage{amssymb}
\usepackage{epsfig}
\usepackage{array}
\usepackage{epstopdf}
\usepackage{amsthm}
\usepackage{afterpage}
\usepackage{amsmath,amssymb,amsfonts}
\usepackage{verbatim}
\usepackage{psfrag}
\usepackage{color}
\usepackage{cite}
\usepackage{longtable}
\usepackage{multirow}
\usepackage{setspace}
\usepackage{adjustbox}
\usepackage{epsfig}
\usepackage{epstopdf}
\usepackage{footmisc}
\usepackage{fmtcount}
\usepackage{stackrel}
\usepackage{float}
\usepackage{breqn}
\usepackage{enumerate}
\usepackage{graphicx}
\usepackage{subcaption}
\usepackage{tabularray}
\usepackage{algorithm}
\usepackage{algpseudocode}
\usepackage{dblfloatfix}
\usepackage{flushend}
\usepackage{graphicx}
\usepackage{dblfloatfix}

\usepackage{tabularx,booktabs}

\usepackage[labelformat=simple]{subcaption}

\DeclareCaptionLabelFormat{subcaptionlabel}{\normalfont(\textbf{#2}\normalfont)}
\newcolumntype{b}{X}
\newcolumntype{s}{>{\hsize=.5\hsize}X}
\begin{document}

\title{
LoRa Communication for Agriculture 4.0: Opportunities, Challenges, and Future Directions
 }

\author{Lameya Aldhaheri, Noor Alshehhi, Irfana Ilyas Jameela Manzil, Ruhul Amin Khalil, \IEEEmembership{Member, IEEE}, Shumaila Javaid, \IEEEmembership{Member, IEEE}, Nasir Saeed, \IEEEmembership{Senior Member, IEEE}, and Mohamed-Slim Alouini, \IEEEmembership{Fellow, IEEE}\\
\thanks{Lameya Aldhaheri, Noor Alshehhi, Irfana Ilyas Jameela Manzil, Ruhul Amin Khalil, and Nasir Saeed are with the College of Engineering, United Arab Emirates University,
Al Ain 15551, United Arab Emirates Email: {\{202101798; 202119340; 700038120; ruhulamin; nasir.saeed\}}@uaeu.ac.ae}
\thanks{Shumaila Javaid is with the Department of Control Science and Engineering, College of Electronics and Information Engineering, Tongji University, Shanghai 201804, China Email: shumaila@tongji.edu.cn}
\thanks{Mohamed-Slim Alouini is with the Computer, Electrical and Mathematical Science and Engineering (CEMSE) Division, King Abdullah University of Science And Technology 23955, Saudi Arabia Email: slim.alouini@kaust.edu.sa}
\thanks{Manuscript received \today ; revised ~X, X~X, accepted ~X, X~X.}
\markboth{IEEE INTERNET OF THINGS JOURNAL,~Vol.~X, No.~X, X~X}%
 }

\maketitle
\begin{abstract}
The emerging field of smart agriculture leverages the power of the Internet of Things (IoT) to revolutionize farming practices. This paper investigates the transformative potential of Long Range (LoRa) technology as a key enabler of long-range wireless communication for agricultural IoT systems. By critically reviewing existing literature, we identify a lacuna in research specifically focused on LoRa's prospects and challenges from a communication perspective in smart agriculture.
We delve into the details of LoRa-based agricultural networks, encompassing network architecture design, Physical Layer (PHY) considerations tailored to the agricultural environment, and the development of channel modeling techniques that account for unique soil characteristics. The paper further explores relaying and routing mechanisms that address the challenges of extending network coverage and optimizing data transmission in vast agricultural landscapes.
Transitioning to practical considerations, we discuss sensor deployment strategies and energy management techniques, providing valuable insights for real-world deployments. A comparative analysis of LoRa with other prevalent wireless communication technologies employed in agricultural IoT applications highlights its strengths and weaknesses within this specific context.
Furthermore, the paper outlines several future research directions to leverage the potential of LoRa-based agriculture 4.0. These include advancements in channel modeling for heterogeneous farming environments, developing novel relay routing algorithms,  integrating emerging sensor technologies like hyper-spectral imaging and drone-based sensing, on-device Artificial Intelligence (AI) models, and sustainable solutions. This survey can serve as a cornerstone for researchers, technologists, and practitioners seeking to understand, implement, and propel smart agriculture initiatives utilizing LoRa technology.
\end{abstract}

\begin{IEEEkeywords}
Agriculture IoT, LoRa,  smart farming, channel modeling, relaying, routing
\end{IEEEkeywords}

\section{Introduction}
The Internet of Things (IoT) is a pivotal force shaping the contemporary and future landscape across diverse industries, influencing everyday life with unprecedented intelligence. The profound impact of IoT is evident in its rapid integration into various sectors, including agriculture 4.0 \cite{gyamfi2024agricultural,abbasi2022digitization}. The agriculture sector is undergoing a comprehensive transformation driven by the rapid adoption of IoT technologies, revolutionizing traditional agricultural practices, and has also imparted newfound efficiency, financial viability, and sustainability to farming endeavors \cite{rose2018agriculture}.

The integration of IoT in agriculture has brought significant advancements, merging innovation with traditional practices. Numerous IoT applications, such as those supporting irrigation, water monitoring, soil analysis, and pesticide management, play a vital role in optimizing farming processes \cite{8793165}. Enabled by the collaboration of information technology and IoT, these applications form a crucial link that enhances farming efficiency. Additionally, the use of remote systems, wireless communication, and monitoring sensors highlights the diverse and multifaceted nature of IoT-driven agricultural solutions. This integration broadens the range of agricultural practices and promotes the development of smart farming techniques. By leveraging IoT, agriculture can proactively tackle challenges, providing practical solutions that drive growth and guide the sector towards sustainability and advanced technological progress \cite{9823520}.

\begin{figure}[htp!]
\begin{center}
\includegraphics[width=1\columnwidth]{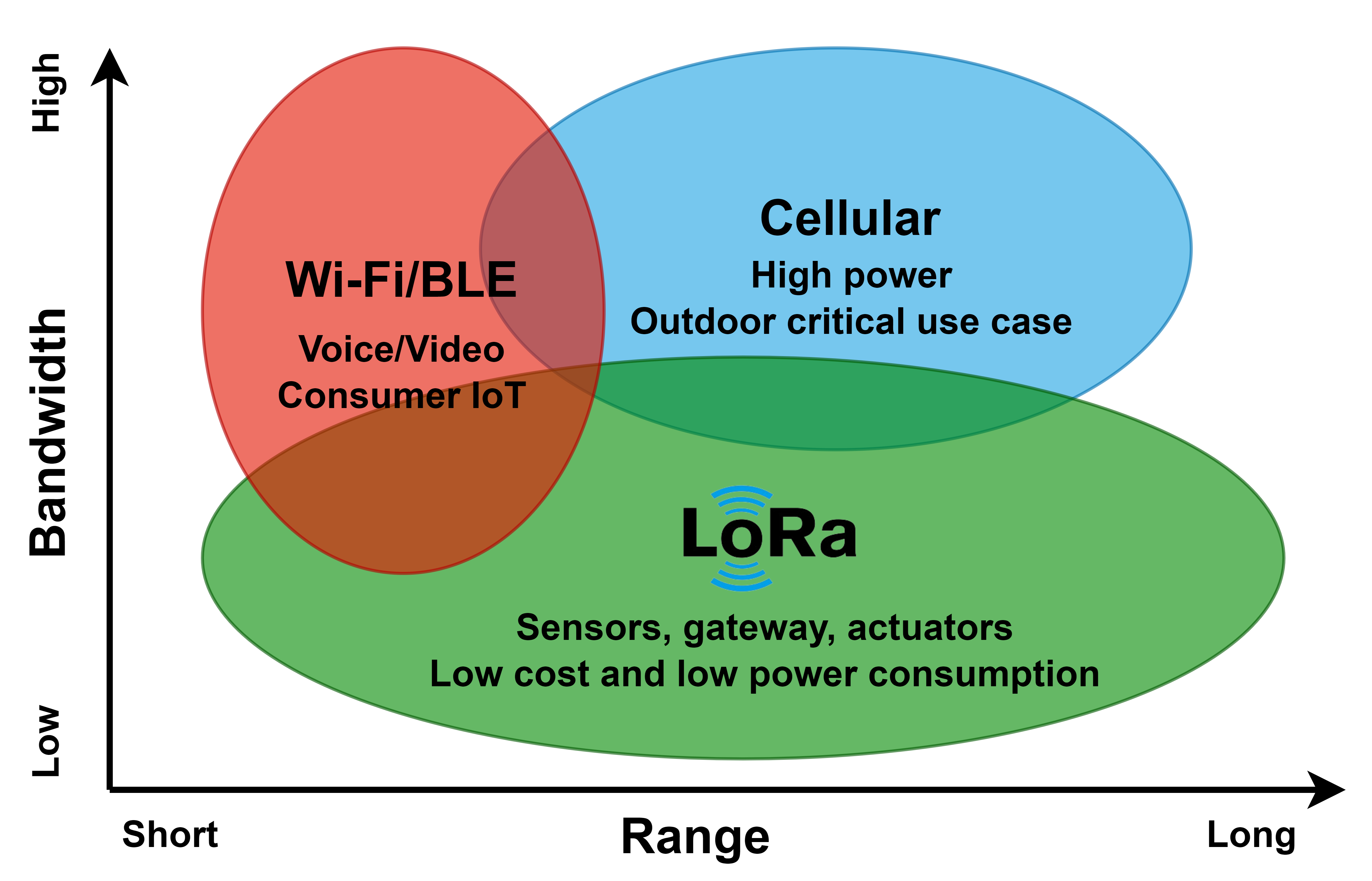}
\caption{LoRa vs. other communication technologies.}\label{LoRa}
\end{center}
\end{figure}
\begin{figure*}[t!]
\begin{center}
\includegraphics[width=1\textwidth]{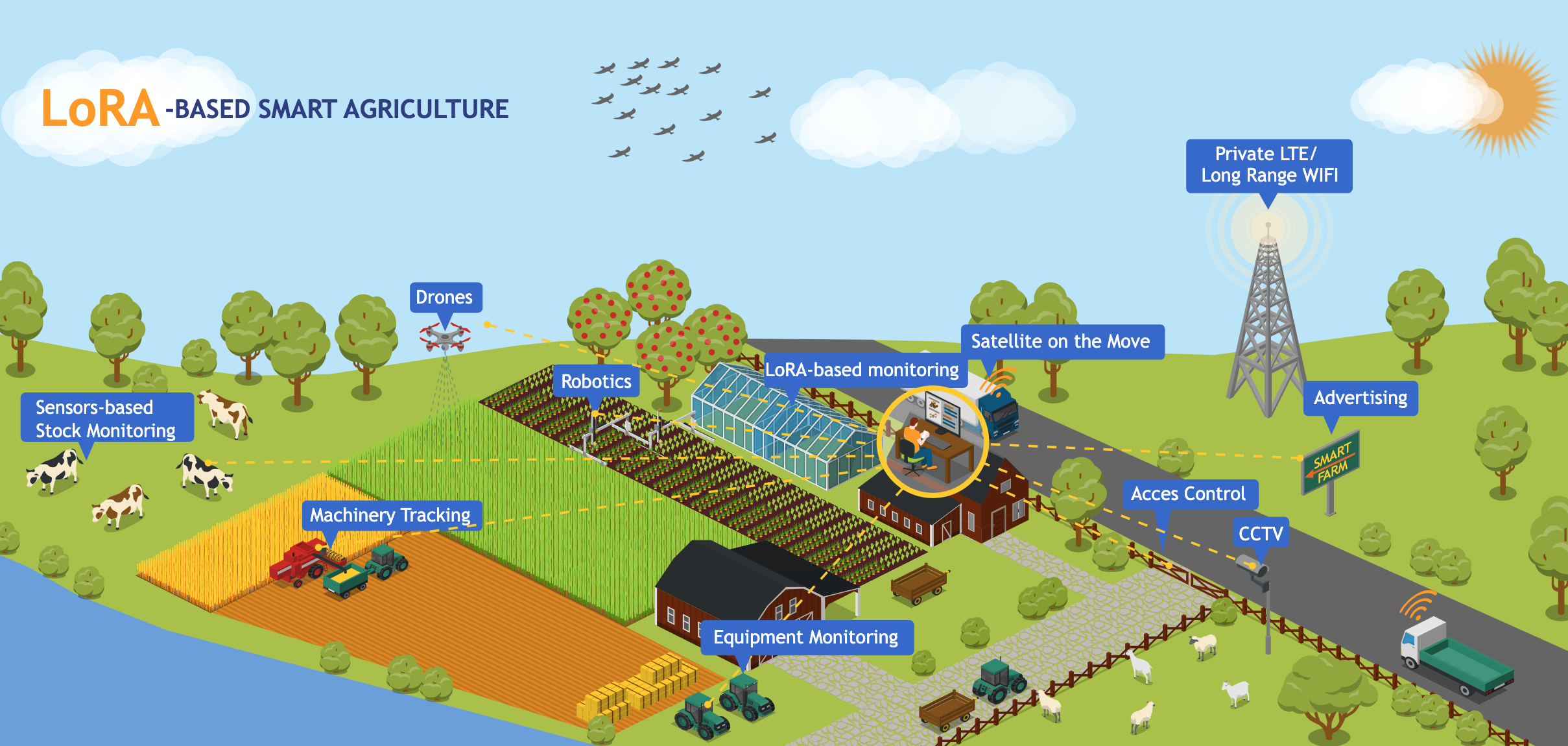}
\caption{LoRa-based smart agriculture farming network.}\label{smartagriculture}
\end{center}
\end{figure*}

The establishment of agriculture IoT necessitates the utilization of various wireless communication technologies, which facilitate enhanced monitoring and control systems across diverse applications \cite{ahmed2018internet,quy2022iot,cambra2017iot,farooq2020role,ruan2019agriculture,xu2022review}. The commonly utilized wireless communication technologies in agricultural IoT are ZigBee, Bluetooth, Sigfox, Wireless Fidelity (WiFi), and Long Range (LoRa) \cite{khanna2019evolution}. These technologies exhibit distinct variations in data communication range and power efficiency, making certain technologies more optimal for short-range monitoring, while others are better suited for long-range monitoring and control applications \cite{9609421,khalil2021deep}. Specifically, ZigBee is effective at short-distance communication but lacks efficacy in long-distance transmission. SigFox demonstrates high power efficiency, though it operates at a comparatively low data rate. Conversely, LoRa is engineered for long-range monitoring and management of agricultural parameters, offering clear performance advantages in these applications \cite{10179693,augustin2016study}. Hence, LoRa has emerged as a pivotal and innovative technology that shapes today's interconnected systems for low-power and long-range communication, as illustrated in Fig.~\ref{LoRa}. Noteworthy for its capacity to facilitate long-range wireless communication, LoRa technology extends beyond mere connectivity, incorporating attributes such as low power consumption and secure data transmission. A key factor enhancing LoRa's performance is its use of Chirp Spread Spectrum (CSS) modulation, which minimizes signal interference and noise, thereby improving the reliability of data exchange \cite{8765617}. This combination of long-range communication and robust data transmission makes LoRa a highly efficient solution for various IoT applications.

Numerous studies have explored the integration of LoRa technology across various sectors, including smart farms, healthcare, and homes. A notable example is presented in \cite{9163432}, where the authors introduced a modified LoRa architecture utilizing the energy-efficient LoRa@FIIT protocol to optimize communication parameters. This optimization is particularly crucial for healthcare devices, where the accurate and reliable transmission of vital data, followed by acknowledgment, is essential. Additionally, LoRa’s capacity to support long-distance communication, enabled through the deployment of base stations and gateways (GWs), allows coverage over several kilometers, making it especially valuable in smart city infrastructures. These features further solidify LoRa as a preferred choice for a wide range of IoT applications. Fig. \ref{smartagriculture} illustrates a LoRa-based smart agriculture setup with different entities.

Furthermore, LoRa boasts an enhanced link budget compared to alternative technologies \cite{8580479, miles2020study}. For example, in \cite{changqing2018internet}, the authors have successfully implemented LoRa wireless networks in greenhouses, utilizing solar-powered sensor collection points for flexible positioning within the greenhouse environment. In \cite{chen2020aIoT}, Chen et al. demonstrated the effectiveness of LoRa modules in transmitting data for pest detection systems to central servers, showcasing the adaptability of LoRa technology in enhancing agricultural monitoring and management practices.  In \cite{tacskin2020long, codeluppi2020lorafarm, adami2020monitoring}, the authors have proposed IoT platforms based on Long Range Wide Area Network (LoRaWAN) architecture due to its simplicity, modularity, and broad deployment possibilities.

The unique feature of LoRa, known as Spreading Factor (SF), allows for customizable trade-offs between coverage area, data rates, and radio packet size, making it an ideal choice for applications like acoustic fish telemetry, as highlighted in \cite{hassan2019internet}. Obstructions and structures within that area primarily influence the effective range of LoRa in a given environment. In contrast to many wireless systems employing frequency shift keying for power conservation, LoRa distinguishes itself by utilizing CSS modulation. This approach proves particularly advantageous for applications in long-distance communication sectors, enabling efficient long-distance communication with minimal power consumption as presented in \cite{9243449}. This strategic use of CSS optimizes LoRa technology for diverse communication scenarios.

Due to the aforementioned advantages of LoRa, in this paper, we present the state-of-the-art LoRa-based IoT applications in agriculture, which serve as a scholarly consolidation of current knowledge in the field, providing a valuable resource for researchers and industry professionals. By examining the advancements and applications of LoRa technology in agriculture, the paper endeavors to bridge the gap between theoretical insights and practical implementations. Researchers will explore the latest trends, methodologies, and challenges in depth. At the same time, industry experts can leverage this information to make informed decisions for successfully integrating LoRa-based solutions into agricultural IoT systems.
This review's synthesis of academic rigor and practical relevance establishes it as an essential resource for those seeking a comprehensive understanding of the potential and challenges associated with LoRa technology in agricultural IoT.
This article presents an overview of LoRa technology for Agriculture IoT and analyzes its application in various agricultural situations. The study examines numerous aspects of LoRa, highlighting its significance and advantages in precision farming and farm monitoring, practical implementation, channel modeling, networking, and prototypes.

\newcolumntype{C}{>{\arraybackslash}X} 
\setlength{\extrarowheight}{1pt}
\begin{table*} [h!]
 \caption{Comparison of this work with other surveys.}
\label{table0001}
 \begin{tabular}
{|p{2.0cm}|p{2cm}|p{10cm}|p{2.5cm}| }
\hline
\hline
\textbf{References} & \textbf{Year}
& \textbf{Brief Description} & \textbf{Contributions Towards Smart Agriculture}\\
\hline
\cite{raza2017low}  & 2017 & Discusses various designs and methods to use in the implementation of LPWAN with low-power devices for a wide-area coverage & $\times$ \\
\hline
\cite{sinha2017survey} & 2017 & Presents a comparative analysis of  LoRa and Narrowband (NB-IoT) for MAC protocols & $\times$ \\
\hline
\cite{migabo2017comparative} & 2017 & Provides a comparative analysis of LoRa and NB-IoT along the current research trends in terms of coverage, reliability, and energy efficiency & $\times$ \\
\hline
\cite{ayoub2018internet} & 2018 & Discusses different mobility factors and their connectivity in LPWAN  applications & $\times$  \\
\hline
\cite{sundaram2019survey} & 2019 & Presents issues and possible solutions of LoRa in terms of networking & $\times$ \\
\hline
\cite{erturk2019survey} & 2019 & Discusses LoRaWAN technology along with its challenges in terms of architecture & $\times$ \\
\hline
\cite{osorio2020routing} & 2020 & Provides a detailed summary of LoRa-based routing protocols and multi-hop communication networks & $\times$ \\
\hline
\cite{kufakunesu2020survey} & 2020 &  Provides a detailed review of the current trends and techniques used of data rate optimization for LoRaWAN networks & $\times$ \\
\hline
\cite{noura2020lorawan} & 2020  & Provides an overview of different security and privacy vulnerabilities for LoRa-based networks & $\times$ \\
\hline
\cite{staikopoulos2020image} & 2020 & Highlights LoRa-based schemes available in literature in terms of image transformations & $\times$ \\
\hline
\cite{shahjalal2020overview} & 2020 &  Presents an overview of the use of AI for LoRa-based smart home monitoring & $\times$ \\
\hline
\cite{marais2020survey} & 2020 & Provides an overview of the recent trends for LoRa, focusing on data trafficking issues & $\times$ \\
\hline
\cite{boquet2021lr} & 2021 & Highlights the importance of LR-FHSS for LoRa-based networks & $\times$ \\
\hline
\cite{benkahla2021review} & 2021 & Provides a detailed review of and experimental assessments of the current ADR for LoRaWAN networks & $\times$ \\
\hline
\cite{ghazali2021systematic} & 2021 & Presents an overview of the current research trends in UAV-based LoRa-communication in real-time scenarios & $\times$ \\
\hline
\cite{jouhari2023survey} & 2023 & Highlights current security protocols, interference reduction, and network scalability developments to solve issues in managing dense IoT networks using LoRaWAN & $\times$ \\
\hline
\cite{9993728} & 2023 & Discusses LoRa-based smart agriculture, focusing mainly on applications & $\checkmark$ \\
\hline
This survey paper & 2024 & \begin{itemize}
    \item Presents a detailed overview of LoRa-based smart agriculture IoT architecture.
    \item Discusses various issues of LoRa-based agriculture IoT, including Network/PHY layers, data acquisition, actuation control, and power management.
    \item Introduces channel modeling techniques and relay routing algorithms tailored for LoRa-based smart agriculture IoT.
    \item Provides an overview of practical implementations of LoRa-based agriculture IoT systems.
    \item Compares LoRa with other wireless communication technologies for smart agriculture IoT applications.
    \item Identifies several future research directions, such as standardization, advanced channel models, multi-sensor integration, and positioning techniques specific to LoRa-based smart agriculture IoT.
\end{itemize} & $\checkmark$   \\
\hline
\hline
\end{tabular}
\end{table*}

\subsection{Related Survey Papers}

LoRa has become indispensable in IoT due to its long-range wireless communication and low power consumption, transforming industries like smart cities \cite{althobaiti2023robust}, industrial automation \cite{khalil2020network}, and smart agriculture. Research has emerged in various areas, particularly collision prevention and interrupt management in LoRa networks. For instance, the authors in \cite{noura2020lorawan} and \cite{raza2017low} conducted a comprehensive study on LoRaWAN architecture, identifying components and classifying attacks and countermeasures. In \cite{sinha2017survey, migabo2017comparative, erturk2019survey}, the authors compared various Low-Power Wide-Area Network (LPWAN) technologies, emphasizing LoRa's adoption. Osorio et al. \cite{osorio2020routing} summarized LoRa routing protocols but did not delve into its impact on PHY layers and energy consumption. Studies also explored LoRa's application in imaging-based communication \cite{staikopoulos2020image} and Machine Learning (ML) for smart-home monitoring \cite{shahjalal2020overview}. Other works, such as \cite{sundaram2019survey} and \cite{marais2020survey}, addressed LoRa network issues and traffic issues in LoRaWAN use cases.
The authors in \cite{boquet2021lr} detailed the applications of the Long Range-Frequency Hopping Spread Spectrum (LR-FHSS) as an extension of the LoRa PHY layer. In contrast, Ayoub et al. \cite{ayoub2018internet} compared LPWAN technologies, focusing on mobile device management. The authors in \cite{ghazali2021systematic} reviewed Unmanned Aerial Vehicle (UAV) based LoRa communication networks, while Adaptive Data Rate (ADR) mechanisms were investigated in \cite{benkahla2021review} and \cite{kufakunesu2020survey}. The authors in \cite{gkotsiopoulos2021performance} identified critical factors for robust LoRa services.
Recently, Jouhari et al. \cite{jouhari2023survey} provided a detailed survey on LoRa technology, including its PHY and Media Access Control (MAC) layer protocols for massive IoT networks. In \cite{sun2022recent}, the authors investigated the major networking challenges arising from the growing demand for diverse IoT applications, which has led to the development of key technologies, including LoRa. They presented a systematic review of LoRa's performance, communication improvements, security vulnerabilities, and various applications, highlighting key advancements in these areas. The authors in \cite{marquez2023understanding} provided a comprehensive review of various localization techniques used in LoRa technology, with a focus on their performance metrics and challenges. It presented a comparative analysis of the techniques, highlighting unresolved issues and offering insights into selecting the most suitable localization system for specific scenarios. In another work \cite{alipio2023current}, the authors studied the testing and evaluation methodologies used for LoRa and LoRaWAN technologies in various IoT applications. The authors classified and analyzed existing studies based on test parameters, architectures, and performance evaluation methods to identify challenges in current testing practices.

Although the surveys mentioned above are comprehensive, none specifically address the investigation of the physical and network layer issues for LoRa technology in smart agriculture IoT. This highlights potential gaps in the existing literature and indicates the need for further research and investigation into applying LoRa in smart agriculture. A detailed comparison of our review article with earlier surveys is provided in Table \ref{table0001}.

\subsection{Motivation of this Survey}
Integrating LoRa technology in smart agriculture has received limited attention in existing surveys and studies, underscoring its potential to enhance farming efficiency and sustainability. For example, \cite{9993728} offers an overview of LoRa's trends and prospects in smart agriculture, emphasizing agricultural applications while ignoring the technical aspects, such as layer-by-layer discussion of the prospects and challenges.
In contrast, our paper goes beyond previous surveys by exploring advancements and applications of LoRa technology in agriculture while identifying gaps and challenges. Unlike prior works, we delve into practical implementation challenges, PHY layer channel modeling specifics, relay and routing protocols, AI  and ML integration for data analysis, and practical implementations, offering a comprehensive view of LoRa's application in agriculture 4.0. By emphasizing the need for standardization, energy efficiency, on-device AI models, localization and positioning, and multi-sensor integration, our paper suggests future research directions to fully leverage LoRa's capabilities, serving as a valuable resource for researchers, technologists, and practitioners striving to advance Agriculture 4.0.

\subsection{Contributions}
In this section, we outline the primary contributions of our survey paper, which focuses on LoRa technology and its applications within smart agriculture IoT. Our contributions encompass:

\begin{itemize}
\item Firstly, we elucidate the potential framework of LoRa technology within smart agriculture, addressing application, network, and PHY challenges.

\item Subsequently, we provide a comprehensive exposition on channel modeling for LoRa in agricultural applications, encompassing both signal propagation in soil and from soil to above-ground.

\item Following that, we emphasize the significance of network layer issues, including relaying and routing for Underground to AboveGround (UG2AG) communication in agriculture IoT networks.

\item Then, we identify areas within the domain of smart agriculture where research on applying LoRa technologies is currently lacking. This involves examining various aspects, including network architecture design, PHY considerations, and path modeling techniques tailored to agricultural environments' unique characteristics and requirements.

\item Next, we delve into the practical considerations of implementing LoRa-based agricultural IoT systems for real-world applications. This encompasses sensor deployment strategies, energy management techniques to prolong device lifespan, and the seamless integration of LoRa technology into existing agricultural infrastructure.

\item Furthermore, we conduct a detailed comparative analysis of LoRa with other prevalent wireless communication technologies commonly utilized in agricultural IoT applications, such as WiFi, Radio-Frequency Identification (RFID), and Bluetooth. Through this comparison, we evaluate these technologies' advantages, limitations, and suitability for various agricultural use cases.

\item Finally, the paper concludes by outlining potential avenues for future research and development in the realm of LoRa-based smart agriculture and IoT. This includes discussing topics such as standardization efforts to ensure interoperability, energy-efficient communication protocols, effective multi-sensor integration strategies, and leveraging emerging technologies like AI and ML to enhance the capabilities of LoRa-based agricultural systems.
\end{itemize}

\subsection{Organization}
The rest of the survey paper is organized as follows. Section \ref{loraforAgriculture} discusses the specifics of implementing LoRa-based agriculture networks, covering network architecture, including application, network, and PHY.  Section \ref{channelmodeling1} presents an overview of channel modeling for LoRa in smart agricultural IoT in the context of Underground-to-Underground (UG2UG) and UG2AG links. Section \ref{relayrouting} presents relaying and routing mechanisms for LoRa-based smart agriculture IoT networks. The practical implementation considerations for agriculture IoT are presented in Section \ref{practicalimplementation}. Following this, we present a comparative analysis of LoRa with other wireless communication technologies commonly used in agriculture IoT in Section \ref{comparison1}. Section \ref{AI-LoRAWAN} presents a detailed discussion of existing studies on AI-enabled LoRaWAN. Finally, in Section \ref{futuredirections}, we outline future research directions to guide further exploration and development in this field, followed by concluding remarks in Section \ref{conclusions}.

\section{LoRa-based Agriculture IoT Network}\label{loraforAgriculture}
\begin{figure}[h!]
\begin{center}
\includegraphics[width=1\columnwidth]{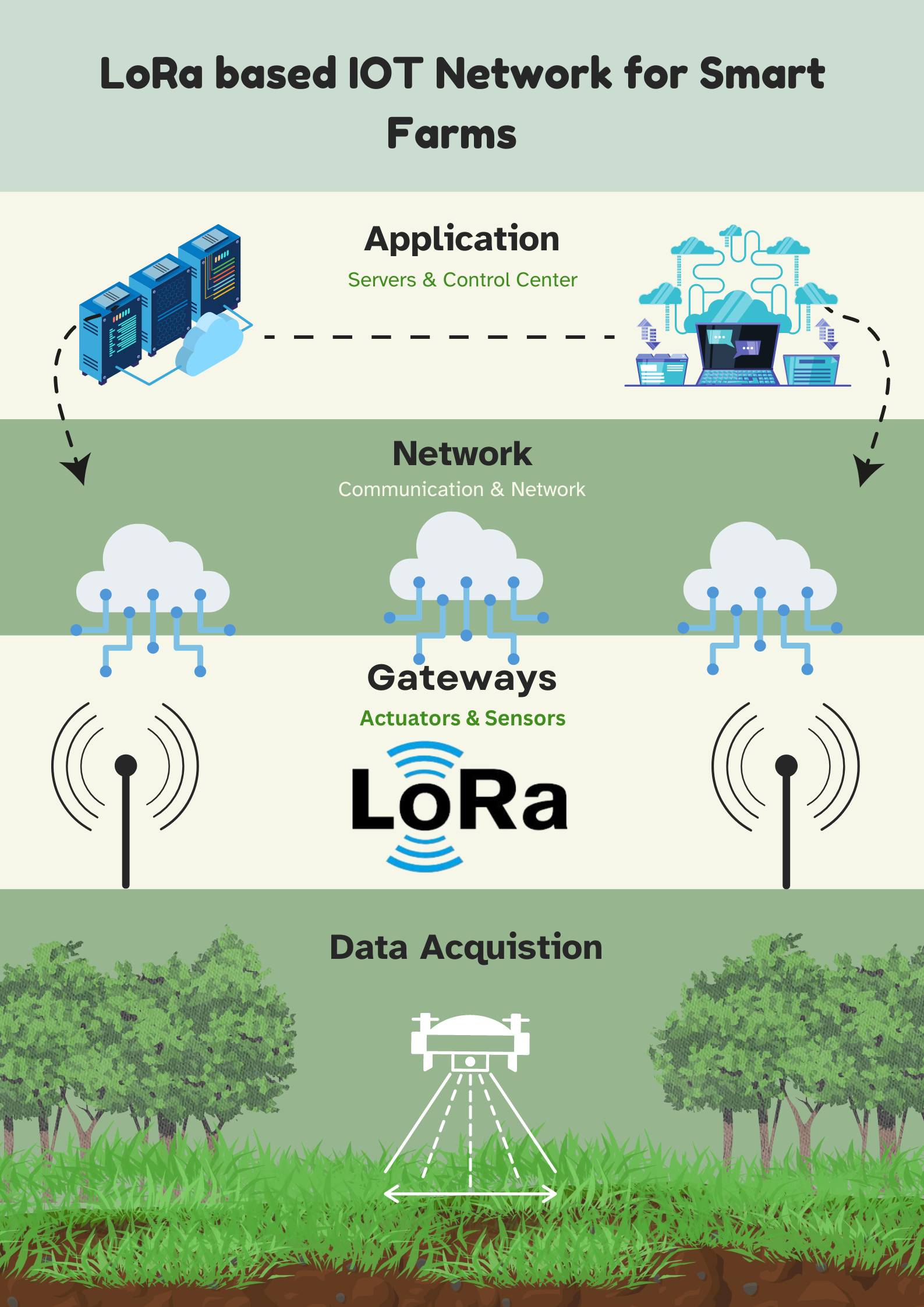}
\caption{LoRa-based IoT network architecture for smart agriculture.}\label{architecture}
\end{center}
\end{figure}

The LoRa-based agriculture IoT network represents a significant advancement in modern technology explicitly tailored for the expansive landscapes of smart farms. The architectural design prioritizes efficient, long-range communication aimed at conserving power while facilitating the transmission of critical data across various components within the farm \cite{kuaban2019architectural,swain2021cost}.
Fig. \ref{architecture} illustrates the overall layer-by-layer architecture of the LoRa-based IoT network in an agriculture 4.0 setup, consisting of the application layer, network layer, and PHY \cite{chanwattanapong2021lora}.
The application layer functions as a bridge, interpreting the data and facilitating real-time decisions. It ensures seamless integration of information, while the network layer acts as a pathway for data transfer from the agricultural fields.
Lastly, the PHY instruments capture the farm's vital signs—monitoring soil moisture levels, assessing crop health, tracking climatic conditions, and communicating the information via LoRa technology. This architectural framework exemplifies the transformative potential of IoT in revolutionizing agricultural practices.
Farmers can monitor their farm's requirements and proactively anticipate them through the seamless integration of these architectural components, ensuring optimal conditions for each seed to realize its fullest potential. In the following, we discuss each of these layers in detail.

\subsection{Application Layer}
The application layer functions as the strategic command center within the LoRa-based Agriculture IoT Network, residing in the cloud or local servers. Here, the convergence of data from various farm sources occurs, transforming raw datasets into actionable insights. Visualize a sophisticated dashboard adorned with graphs and alerts, each element contributing to the comprehensive understanding of the farm's status.
Within servers housed within this layer, data undergoes rigorous processing, leveraging algorithms and ML techniques to forecast, strategically, and respond effectively to agricultural demands \cite{suji2022efficient,akhter2022precision,shaikh2022towards}. For example, these servers may analyze historical crop yields alongside meteorological data to offer optimized planting schedules.
Moreover, the application layer facilitates remote farm management, empowering farmers to oversee and regulate every facet of their operations, even when physically distant. From adjusting irrigation schedules to determining optimal harvesting times, this layer affords farmers a level of foresight akin to possessing a figurative crystal ball into the future of their crops.

\subsection{Network Layer}
\begin{figure}[h!]
\begin{center}
\includegraphics[width=0.5\textwidth]{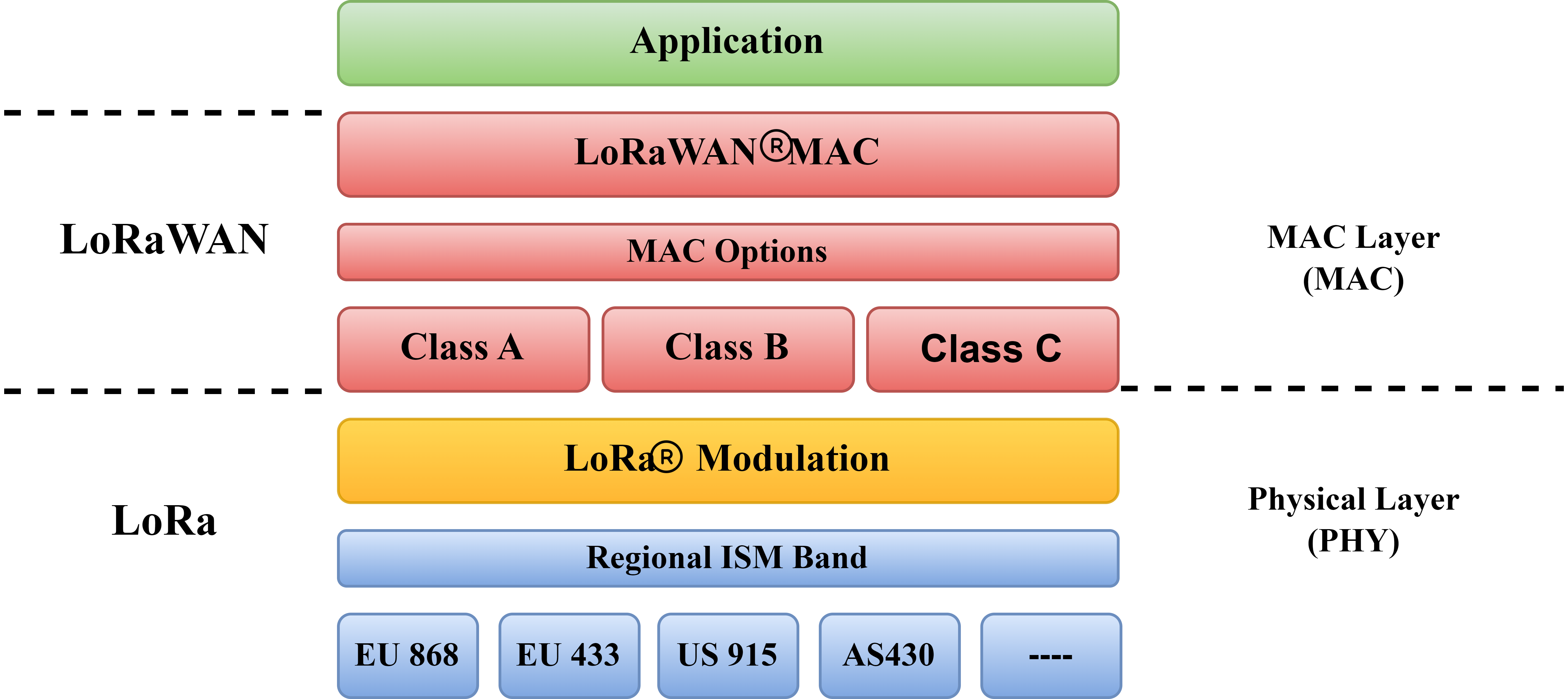}
\caption{MAC and PHY layers of LoRa.}\label{macphylayers}
\end{center}
\end{figure}

The network layer architecture for LoRa in smart agriculture assumes a pivotal role in bridging the connectivity divide between physical IoT devices distributed across the agricultural landscape and the digital realm, where data assumes actionable significance \cite{jouhari2023survey,yascaribay2022performance,citoni2019internet}. This architectural framework encapsulates fundamental functionalities as shown in Fig. \ref{macphylayers}, from GWs management to device state monitoring and application data processing. Key constituents include nodes, GWs, routers, brokers, Network Servers (NSs), and handlers, each tasked with distinct roles spanning data transmission to encryption, ensuring seamless data flow and processing within the network. GWs transcend mere transmission nodes, playing integral roles in optimizing network traffic and preserving message integrity through duplication mechanisms and metadata oversight. Leveraging these GWs, the network negotiates the complexities inherent in transmitting non-IP protocol data, such as that facilitated by LoRaWAN, across the Internet for subsequent application processing \cite{swain2021lora,arshad2022implementation}. This decentralized system architecture empowers users to engage with the global network by establishing their segments, thus bolstering the scalability and accessibility of smart agricultural solutions. Brokers and handlers act as facilitators, ensuring the seamless transmission and interpretation of data and effectuating the conversion of raw binary inputs into actionable insights. Consequently, smart agriculture practices are more efficient and responsive to real-time environmental conditions \cite{9993728,yascaribay2022performance,sharofidinov2020agriculture}. This sophisticated network infrastructure, underpinned by LoRaWAN technology, emerges as a linchpin for advancing smart agricultural paradigms, offering scalability, security, and adaptability to accommodate the evolving requisites of modern farming. A core tenet of this architecture is its flexibility in deployment, manifesting in options ranging from seamless application integration via platforms like The Things Stack Sandbox \cite{thethingnet2023} to establishing private networks safeguarding data within localized environments. Such adaptability furnishes solutions tailored to diverse agricultural exigencies, from rudimentary monitoring systems to intricate, data-intensive operations necessitating precision control and analysis. Additionally, the infrastructure lends support to hybrid models, amalgamating the benefits of private networks with those of broader community networks, thereby exemplifying a forward-looking approach to IoT deployment in agriculture.

Within the network layer of LoRa-based smart agriculture systems, protocols such as Message Queuing Telemetry Transport (MQTT) and Constrained Application Protocol (CoAP) assume prominence. MQTT is favored for its lightweight and efficient messaging capabilities, ideal for remote monitoring and control scenarios \cite{ayushjain2023}. At the same time, CoAP is tailored for simple, resource-constrained devices, offering web-like interactions with minimal overhead, rendering both protocols well-suited for the data-intensive yet resource-limited landscape of smart agriculture.

Besides, Device Address (DevAddr) extraction emerges as a critical process in the network layer of LoRa-based systems, wherein each packet's DevAddr serves as a conduit for routing messages to their intended destinations. This 32-bit non-unique address facilitates device identification and management within the network \cite{thethingnet2023}. Given the non-uniqueness of these addresses, the network employs a mapping system, often facilitated through a discovery service, to correlate each message with the appropriate device and application. This system adeptly manages messages even within densely populated networks, ensuring the accurate collection and processing of data from myriad agricultural sensors and devices, thereby enabling meticulous monitoring and management of farm conditions.
Furthermore, the network layer of LoRa-based systems is significantly influenced by the LoRaWAN protocol, which plays a pivotal role in facilitating efficient, long-range communication between End Devices (EDs) and GWs through a managed MAC layer \cite{donmez2018security,centelles2021beyond}. Employing a star-of-stars topology comprising NSs, GWs, and EDs, wherein the EDs communicate with the network exclusively via GWs, enhances network robustness and streamlines connectivity \cite{sornin2015lorawan,magrin2017performance,moysiadis2021extending}. This configuration facilitates data transmission from many sensors and devices deployed across agricultural fields while enabling network commands to be dispatched back to these devices, facilitating adaptive and responsive farming practices.

LoRaWAN supports three device classes to meet the diverse requirements of smart agriculture. Class A devices prioritize energy efficiency, allowing essential communication with minimal power consumption. Class B devices, meanwhile, offer additional communication windows at predetermined intervals, balancing communication frequency with power usage. In contrast, Class C devices are designed for near-continuous data exchange, making them ideal for applications requiring frequent monitoring \cite{sornin2015lorawan,choi2020loradar}.

The release of LoRaWAN v1.0.4 in October 2020 introduces the DeviceTimeReq and DeviceTimeAns MAC commands for Class A devices, enhancing network time synchronization and replacing older commands for enhanced efficiency. Furthermore, this latest version accentuates security enhancements from prior iterations while perpetuating support for roaming management introduced in LoRaWAN 1.1, thus upholding passive and active roaming capabilities. Despite these advancements, devices adhering to older specifications, such as v1.0.2 and v1.0.3, remain prevalent owing to hardware upgrade constraints \cite{loraalliance2016,loraalliance2018}. The ADR feature further refines data transmission by optimizing transmission rate and power usage based on prevailing network conditions, ensuring efficient and effective data transmission from sensors and devices.
The integration of devices into the network is facilitated through LoRaWAN's joining procedures, encompassing Over-the-Air Activation (OTAA) and Activation by Personalization (ABP), which furnish mechanisms for device authentication and secure data transmission \cite{eldefrawy2019formal}. The evolution of LoRaWAN standards, characterized by updates aimed at bolstering security, scalability, and communication efficiency, underscores the protocol's adaptability to the burgeoning demands of smart agriculture, providing a reliable, secure, and scalable framework for deploying IoT technologies in farming. This evolutionary trajectory ensures that the network layer remains adept at supporting various agricultural applications, from essential soil moisture monitoring to intricate, real-time decision-making processes, facilitating more efficient, productive, and sustainable farming practices.

\subsection{Physical Layer}
The PHY of the LoRa-based IoT agriculture system encompasses the tangible components responsible for sensing environmental data and acquiring information directly from the farm's surroundings \cite{wang2020slora}. This layer comprises sensors, actuators, and data acquisition mechanisms that interact with the physical environment to gather real-time data and facilitate actionable insights for farm management, as shown in Fig. \ref{macphylayers}. LoRa uses CSS technology to maintain communication over significantly larger distances than traditional wireless systems, making it highly suitable for IoT applications requiring wide coverage and low power consumption. CSS resilience to interference and capacity for penetrating physical barriers allows for reliable data transmission even in challenging environments. The PHY layer's adaptability, through adjustable parameters such as SF, Bandwidth (BW), and Coding Rate (CR), provides a flexible approach to balancing range, data rate, and power consumption based on application needs. These models can be categorized into empirical, deterministic, and stochastic types for smart agriculture systems and play a crucial role in estimating signal strength based on path-loss calculations \cite{stusek2020accuracy}. For instance, widely used models like the Okumura-Hata and COST 231-Hata models provide mathematical frameworks for predicting path loss, considering factors such as frequency, range, distance between EDs and GWs, and antenna height \cite{joseph2013urban}. Accurate channel modeling improves data transmission efficiency in agricultural applications such as soil monitoring and irrigation control, particularly between UG2AG and UG2UG. UG2UG communication involves analyzing signal propagation through the soil while considering characteristics such as soil type, moisture content, and conductivity \cite{manzil2023performance}  \cite{wang2017primer}. This knowledge contributes to optimizing frequency selection, antenna arrangement, and transmission characteristics, ensuring a reliable link between sensors and GWs. By leveraging these models, planners can optimize network deployment by determining the number and placement of GWs required to ensure adequate coverage and connectivity across agricultural fields.

Furthermore, channel characterization is another essential aspect of the PHY in LoRa-based smart agriculture systems. Establishing the Signal-to-Noise Ratio (SNR) threshold is critical for accurate demodulation of received signals, particularly in the presence of interference \cite{milanovic2007comparison}. Different SNR thresholds are defined for each SF to ensure reliable communication under varying conditions. Also, Reinforcement Learning (RL) can be used to optimize the allocation of SFs in subterranean LoRaWAN and Non-Terrestrial Networks (NTN)  by employing Multi-Agent Dueling Double Deep Q-Network (MAD3QN) and Multi-Agent Advantage Actor-Critic (MAA2C) algorithms \cite{ lin2023energy}. Additionally, the LoRa modulation scheme supports variable data transmission rates using quasi-orthogonal SFs, allowing for adaptive modulation based on deployment requirements. Using Forward Error Correction (FEC) rates helps mitigate errors caused by interference, albeit at the expense of reduced data throughput. Moreover, LoRa's adaptive CSS modulation allows multiple simultaneous transmissions on a single channel, enhancing network efficiency and robustness in agricultural settings.
\begin{itemize}
    \item CSS Modulation: LoRa employs CSS modulation, which encodes data into chirp signals—varying frequency pulses. This technique enhances signal robustness against channel noise and interference, enabling reliable long-range communication. The modulation equation can be represented as
     \begin{equation}\label{signal}
        \begin{aligned}
            S(t)=\cos[2\pi (f_{0}+kt)t]
        \end{aligned}
    \end{equation}
    \noindent where $f_0$ is the initial frequency and $k$ is the frequency rate of change.
    \item ADR: LoRa PHY allows adjusting the data rate according to network conditions and distance between nodes. This adaptability is achieved by altering three key parameters, BW, SF, and CR, optimizing the balance between communication range and data transmission speed.
    \item SF and BW: LoRa's use of SFs from SF7 to SF12 permits a trade-off between data rate and sensitivity, with higher SFs allowing for longer range at lower data rates. The relation between SF and the symbol rate $R_s$ is given by \eqref{SFB}, indicating how BW and SF impact the symbol rate.
    \begin{equation}\label{SFB}
        \begin{aligned}
            R_{s}=\frac{SF+BW+CR}{2^{SF}}
        \end{aligned}
    \end{equation}
        \item Link Budget and Range Extension: The link budget in LoRa, which determines the maximum range, is significantly improved through CSS and the selection of SFs and BW. This allows communication over tens of kilometers in rural areas, making LoRa ideal for applications requiring long-range connectivity with minimal power consumption.
    \item FEC: LoRa incorporates FEC to enhance data integrity over long distances. The CR defines the redundancy level added to the payload, thereby improving the resilience of the transmitted data against errors.
\end{itemize}

The PHY packet structure of LoRa includes essential components such as a preamble, optional header, data payload, and Cyclic Redundancy Check (CRC) field. The preamble synchronizes the receiver with the transmitter and can vary in length depending on configuration settings. The optional header contains information crucial for packet decoding and can be activated or deactivated based on the header mode selected. In explicit mode, the payload length, FEC coding rate, and header CRC are included, while in implicit mode, these are omitted to reduce airtime. This flexibility in packet structure allows LoRa systems to adapt to varying application requirements and environmental conditions, making it well-suited for the diverse needs of smart agriculture, where reliable and efficient communication is essential for optimizing farming practices and maximizing yields. Besides channel modeling, the following factors are critical in designing an optimal LoRa-based agriculture IoT network.

\subsubsection{Sensor Deployment}
Sensor deployment is a critical aspect of the PHY, involving strategically placing sensors throughout the farm to monitor various environmental parameters \cite{bandyopadhyay2020IoT}. These sensors serve as the farm's "eyes and ears," capturing data on soil moisture, ambient temperature, humidity levels, light intensity, and crop health. By providing real-time feedback, sensors enable farmers to observe minute changes in farm conditions that may not be immediately visible, allowing for timely interventions.

As sensor technology continues to evolve, these systems are becoming even more powerful and precise, offering advanced capabilities that go beyond simple environmental monitoring. Today’s sensors, equipped with features such as multi-spectral imaging, drone-based remote sensing, and chemical analysis of soil nutrients, empower farmers with deeper insights for more informed decision-making. These advancements contribute to the broader concept of precision agriculture, where farmers can utilize hyper-local data to apply resources (e.g., water, fertilizer, pesticides) more efficiently and effectively. For example, soil moisture sensors integrated with IoT frameworks can automate irrigation systems, ensuring optimal water usage and reducing waste \cite{ gs2019smart}.

Furthermore, sensor calibration and precision are crucial when deploying sensing systems to ensure that the data collected is reliable and actionable. Key considerations in sensing include sensor sensitivity, measurement range, and environmental durability, especially in agricultural environments where conditions can vary widely. Advanced sensing technologies, combined with sensor fusion techniques, which aggregate data from multiple sensor types (e.g., thermal, optical, and humidity sensors), can significantly enhance the decision-making process by providing a more holistic and accurate view of farm conditions \cite{barrile2022experimenting, ahmad2022technology}.

Accordingly, by deploying sensors at critical locations across the farm, farmers and agricultural stakeholders can comprehensively understand the farm's conditions and make informed decisions to optimize crop cultivation practices \cite{sharofidinov2020agriculture}. Sensor placement considerations may include factors such as crop type, topography, and irrigation infrastructure, ensuring that sensors are positioned to capture relevant data accurately.

\subsubsection{Data Acquisition}
Data acquisition involves collecting, aggregating, and transmitting sensor data to centralized repositories for analysis and decision-making. In IoT agriculture, data acquisition mechanisms may include wireless communication protocols, such as LoRa, cellular networks, or Wi-Fi, which enable seamless data transmission from sensors to central servers or cloud-based platforms \cite{silva2019low}. Additionally, data acquisition systems may incorporate edge computing capabilities, allowing for preliminary data processing and analysis at the sensor node level to reduce latency and BW requirements. By efficiently acquiring and transmitting data, the PHY enables timely access to actionable insights, empowering farmers to optimize resource allocation, mitigate risks, and enhance crop yields \cite{zorbas2019network}.

\subsubsection{Actuation Control}
Actuation control involves using actuators to perform physical tasks or operations based on sensor data and predefined algorithms. Actuators, such as irrigation valves, pumps, motors, and automated machinery, receive commands from the central control center or edge computing devices and execute actions to regulate environmental conditions, apply treatments, or perform agricultural operations \cite{morais2021versatile}. For example, based on soil moisture sensor readings, actuators can initiate irrigation processes to deliver water to crops at optimal times and rates, ensuring efficient water usage and promoting crop health \cite{singh2022lora}. Actuation control plays a pivotal role in implementing precision agriculture techniques, enabling automated and adaptive management strategies tailored to the specific needs of crops and environmental conditions. IoT agriculture systems can enhance operational efficiency, reduce resource waste, and improve farm productivity by integrating actuation control into the PHY.

\subsubsection{Efficient Power Management}
One of the key advantages of LoRa technology is its efficiency in power consumption, which is critical for IoT deployments in remote agricultural settings where access to power sources may be limited or unreliable. LoRa-enabled sensors and devices are designed to operate on low power, allowing them to function for extended periods, ranging from months to years, on a single battery charge \cite{chanwattanapong2021lora}. This reduces the need for frequent maintenance visits to replace batteries, thereby minimizing operational costs and labor requirements. Additionally, the low power consumption of LoRa devices contributes to sustainability efforts by reducing the ecological footprint associated with battery disposal and replacement. Overall, efficient power management is a fundamental aspect of the network layer, ensuring the longevity and reliability of IoT deployments in agricultural environments.

\begin{figure*}[h]
\begin{center}
\includegraphics[width=0.9\textwidth]{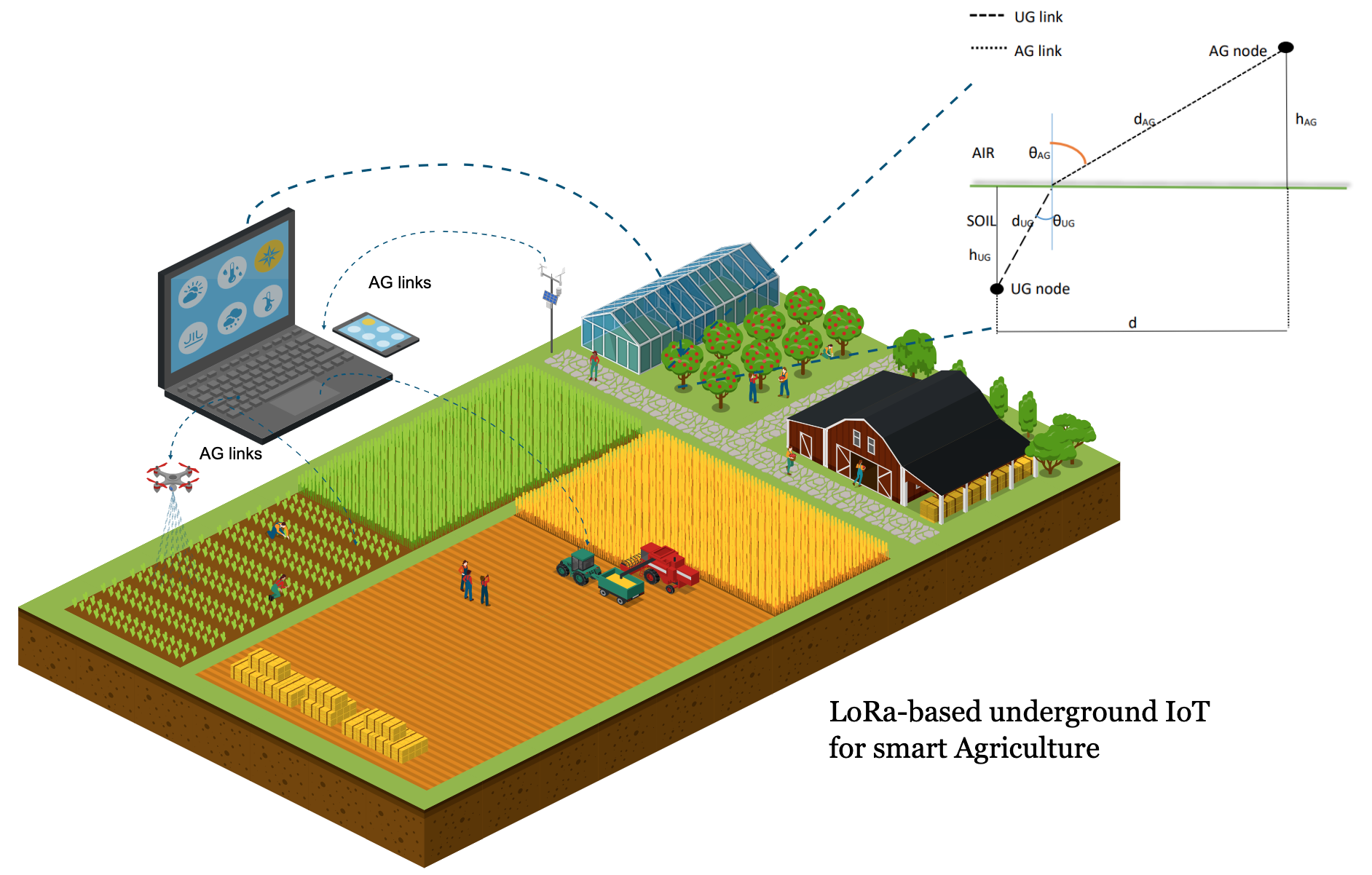}
\caption{Illustration of channel modeling for UG2UG and UG2AG links in soil and air medium.}\label{channelmodeling}
\end{center}
\end{figure*}

\section{Channel Modelling}\label{channelmodeling1}
Channel modeling is a critical aspect of LoRa-based smart agriculture networks, which is fundamental in optimizing communication efficiency and network performance. In the realm of smart agriculture, where remote monitoring and data collection are fundamental, understanding and modeling the propagation characteristics of LoRa signals are essential. Channel modeling enables the prediction of signal behavior in complex agricultural environments, such as orchards or large-scale farms, where factors like vegetation, terrain, and weather conditions can significantly impact signal propagation. Network planners can determine optimal gateway coverage, sensor node placement, and overall network design by developing accurate propagation models tailored to agricultural settings, such as FLog for orchards. FLog is a specialized propagation model for LoRa networks in orchard environments that accounts for the shadowing effects caused by tree canopies, and the ground \cite{yang2023link}. By leveraging the regular spacing and uniform shapes of trees, FLog improves the accuracy of network deployment, ensuring reliable communication for efficient farm management and environmental monitoring. An illustration of channel modeling for UG2AG and AG2UG links in soil and air medium for smart agriculture is depicted in Fig. \ref{channelmodeling}.

Multiple studies have investigated and reported channel models for wireless underground communication channels. In \cite{9361700}, the authors investigated the performance of LoRaWAN from underground to the earth's surface through experimental testing in various soil types, including sand, clay, and gravel. Their results affirm that sandy soil provides the best Received Signal Strength Indicator (RSSI) and SNR performance. In contrast, soil with gravel yields the worst, exhibiting a difference of 10 dBm among the three soil compositions. The test was conducted at a burial depth of 50cm below the surface.
  A similar study was conducted in \cite{hardie2019underground}, employing a 433-MHz LoRa network to evaluate UG2UG and UG2AG wireless communication systems in four different types of soils. However, the transmission distance obtained did not demonstrate any improvement compared to the earlier studies. The highest transmission distance for UG2UG operation was 4-20 m, while UG2AG operation may communicate up to 100-200 m, depending on operating parameters and soil characteristics. The findings indicate that developing 433-MHz LoRa-based UG2AG Wireless Underground Sensor Networks (WUSNs) for agricultural applications is feasible with improvements in power management. However, UG2UG applications remain doubtful without significant improvement in transmission distance.

 In \cite{lei2023experimental}, authors proposed a new UG2AG channel model in LoRaWAN-based UG sensor networks, addressing diffusion loss across multi-layer media for better communication in small-scale complex environments. In \cite{zhao2023optimizing}, a Multi-Agent Reinforcement Learning (MARL) algorithm is introduced to enhance energy efficiency by optimizing transmission configurations considering UG link quality, energy consumption, and packet collisions. Additionally, \cite{zhao2023feasibility} stressed the importance of monitoring UG environments and proposed LoRaWAN-based wireless UG networks. The study highlighted the influence of underground conditions on network performance and the necessity of carefully selecting physical layer parameters. Due to the distinct features of signal propagation in soil, it is necessary to calculate path loss by considering the soil characteristics, including composition and moisture content, as highlighted in the previous studies \cite{9361700} and \cite{manzil2023performance}.
The received power, as formulated by authors in \cite{9361700} within a soil medium, is given by
\begin{equation}
P_{r} = P_{t} + G_{r} + G_{t} - L_{t},
\label{eq:1}
\end{equation}
where, $P_{t}$ is the transmit power, $G_{r}$ and $G_{t}$ are the receiver and transmitter antenna gains, and $L_{t} = L_{AG}+L_{UG}$ is the total path loss. $L_{AG}$ and $L_{UG}$ is the aboveground and underground path loss. The equation for total path loss is given as follows:
\begin{equation}
\ell_{t} = 6.4 + 20 (\log(d)+\log(\beta_{u})) + 8.69\alpha_{u} d,
\label{eq:pathloss}
\end{equation}
where $d$ is the total transmission distance, $\beta_{u}$ is the phase shifting constant, and $\alpha_{u}$ is the attenuation constant that depends on the dielectric properties of soil and is acquired from the Peplinksi model \cite{peplinski1995dielectric}  as given below
\begin{equation}
    \epsilon = \epsilon^'-j\epsilon^{''},
\end{equation}
where $\epsilon^'$ and $\epsilon^{''}$ are the real and imaginary terms of the dielectric constant given as
\begin{equation}
\epsilon^' = 1.15 \left[1 + \frac{D_{b}}{D_{s}} \left(\epsilon_s^{\hat\alpha}\right) + {\theta_v}^{\hat{\beta}} {\epsilon_{w}'}^{\hat\alpha} - \theta_v\right]^{\frac{1}{\hat\alpha}} - 0.68,
\label{eq:realdielectric}
\end{equation}
and
\begin{equation}
{\epsilon^{''}} = \left[\theta_v^{\overline{\beta}}\epsilon_{w}^{{''\hat\alpha}}\right]^{\frac{1}{\hat\alpha}}.
\label{eq:imgdielectric}
\end{equation}
Here, $\theta_v$ is the volumetric moisture content of the mixture,
$D_{b}$ is the bulk density in (g/cm$^3$), $D_{s}$ = 2.66 g/cm$^3$
is the specific density of the solid soil, $\epsilon_{w}'$  and $\epsilon_{w}''$ are the real and imaginary parts of the relative dielectric constant of water, respectively.
$\hat\alpha$ = 0.65,  $\hat\beta = 1.2748-0.519\mu_s-0.152\mu_c $
and $\overline\beta = 1.33797-0.603\mu_s-0.166\mu_c$ are empirically determined constants.
$\mu_s$ and $\mu_c$ indicate the mass ratios of sand and clay. Accordingly, the attenuation constant is
\begin{equation}
\alpha_{u} = \frac{2 \pi C}{\lambda} \sqrt{\frac{\mu \epsilon^'}{2}\left [\sqrt{1+\left(\frac{\epsilon^{''}}{\epsilon^'}\right)^2} -1 \right]},
\end{equation}
and phase shifting constant is
\begin{equation}
\beta_{u} = \frac{2 \pi C}{\lambda} \sqrt{\frac{\mu \epsilon^'}{2}\left [\sqrt{1+\left(\frac{\epsilon^{''}}{\epsilon^'}\right)^2} +1 \right]},
\label{eq:4}
\end{equation}
where $\lambda$ is the wavelength of the signal, and $\mu$ is the magnetic  permeability. Thus, the attenuation and phase-shifting equations indicate that several factors, including operating frequency, soil composition, moisture content, and bulk density, influence the path loss of the signal through the soil.
\begin{figure}[htbp]
\centerline{\includegraphics[width = 1\columnwidth]{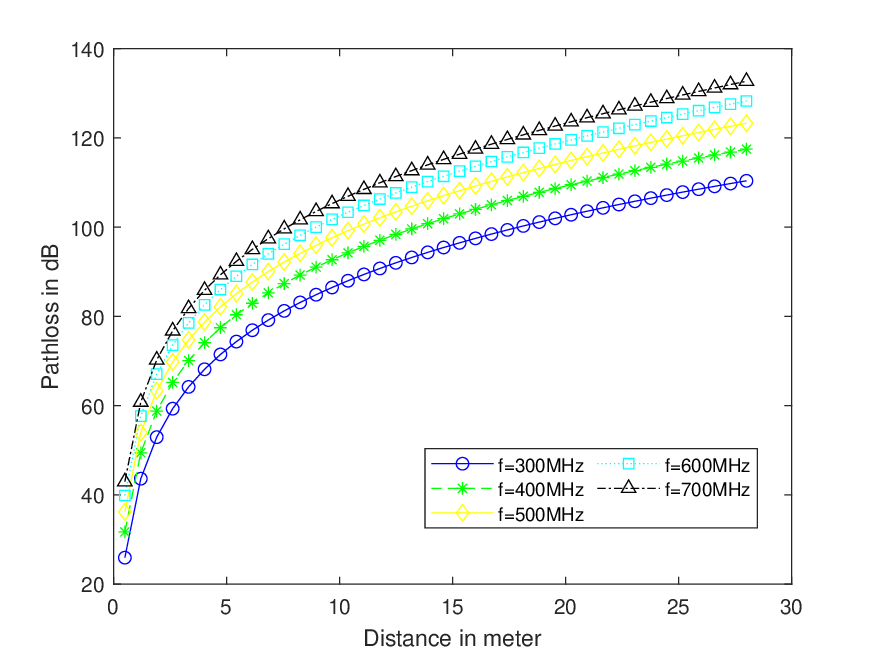}}
\caption{Path loss vs. distance.}
\label{fig:pathlossvsdistance}
\end{figure}

\begin{figure}[htbp]
\centerline{\includegraphics[width = 1\columnwidth]{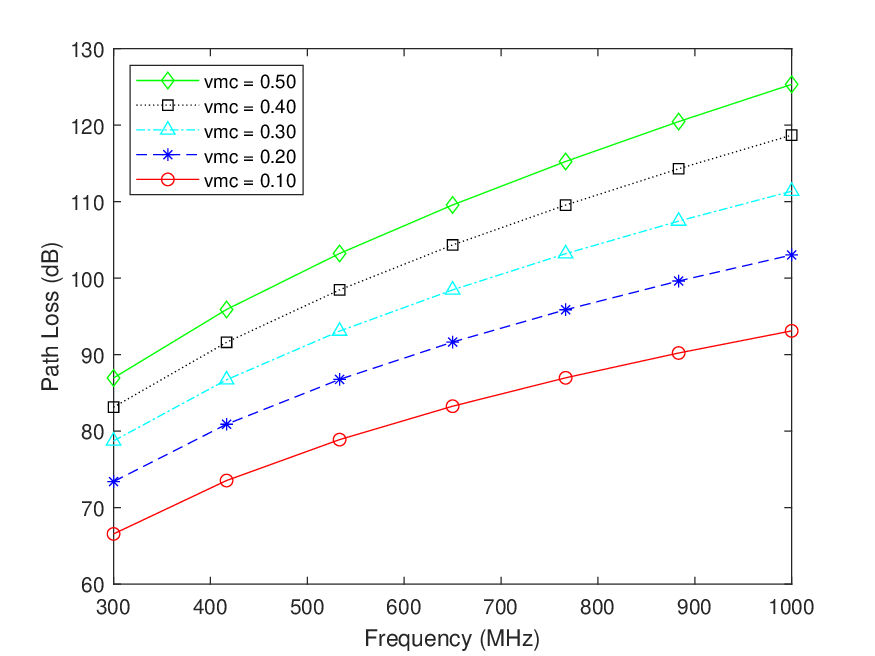}}
\caption{Path loss vs. frequency.}
\label{fig:pathlossvsfrequency}
\end{figure}

To further elaborate, we analyzed these channel characteristics via simulations with the source node buried at a depth of 50cm and the destination node kept above ground to study the influence of propagation distance, frequency, and water content on the path loss. The impact of transmission distance on path loss for variable frequencies (300 MHz - 700 MHz) is shown in Fig. \ref{fig:pathlossvsdistance}. We considered the percentage of the sand particle to be 50\%, clay percentage = 15\%, and bulk density = 1.5 $g/cm^3$, which are taken from \cite{mcvarunchannelmodel}. As expected, the simulations illustrated that the attenuation of the signal continually increases as the distance between the transmitter and receiver grows. This finding is consistent with the theoretical analysis of fundamental principles governing the propagation of radio waves. Furthermore, we have examined the influence of moisture content to validate the impact of soil composition. In Fig. \ref{fig:pathlossvsfrequency}, we demonstrated the influence of Volumetric Water Content (VWC) on path loss spanning from 10\% to 50\%. An elevation in attenuation can be seen when the VWC of the soil increases. To conclude, it is crucial to quantify the attenuation induced by the soil properties, as it is the primary factor affecting wireless underground communication channels.

\section{Relaying and Routing}\label{relayrouting}
Besides channel modeling, relaying and routing are of greater importance in LoRa-based smart agriculture networks, ensuring efficient and reliable communication across vast agricultural landscapes. In smart agriculture, where sensor nodes are deployed over large areas to monitor various parameters such as soil moisture, temperature, and humidity, relaying and routing mechanisms are essential for extending the network coverage and optimizing data transmission. With the proper deployment of relay nodes, the network's reach can be expanded, which can forward data between sensor nodes and GWs by overcoming obstacles such as terrain variations and vegetation obstructing direct communication. On the other hand, routing protocols determine the path data takes through the network, ensuring timely and accurate delivery of information to the appropriate destinations. In the dynamic and challenging environments of agricultural settings, efficient relaying and routing mechanisms are crucial for maintaining connectivity, maximizing network coverage, and enabling seamless data transmission for effective farm management and monitoring.

\begin{figure}[h]
\begin{center}
\includegraphics[width=1\columnwidth]{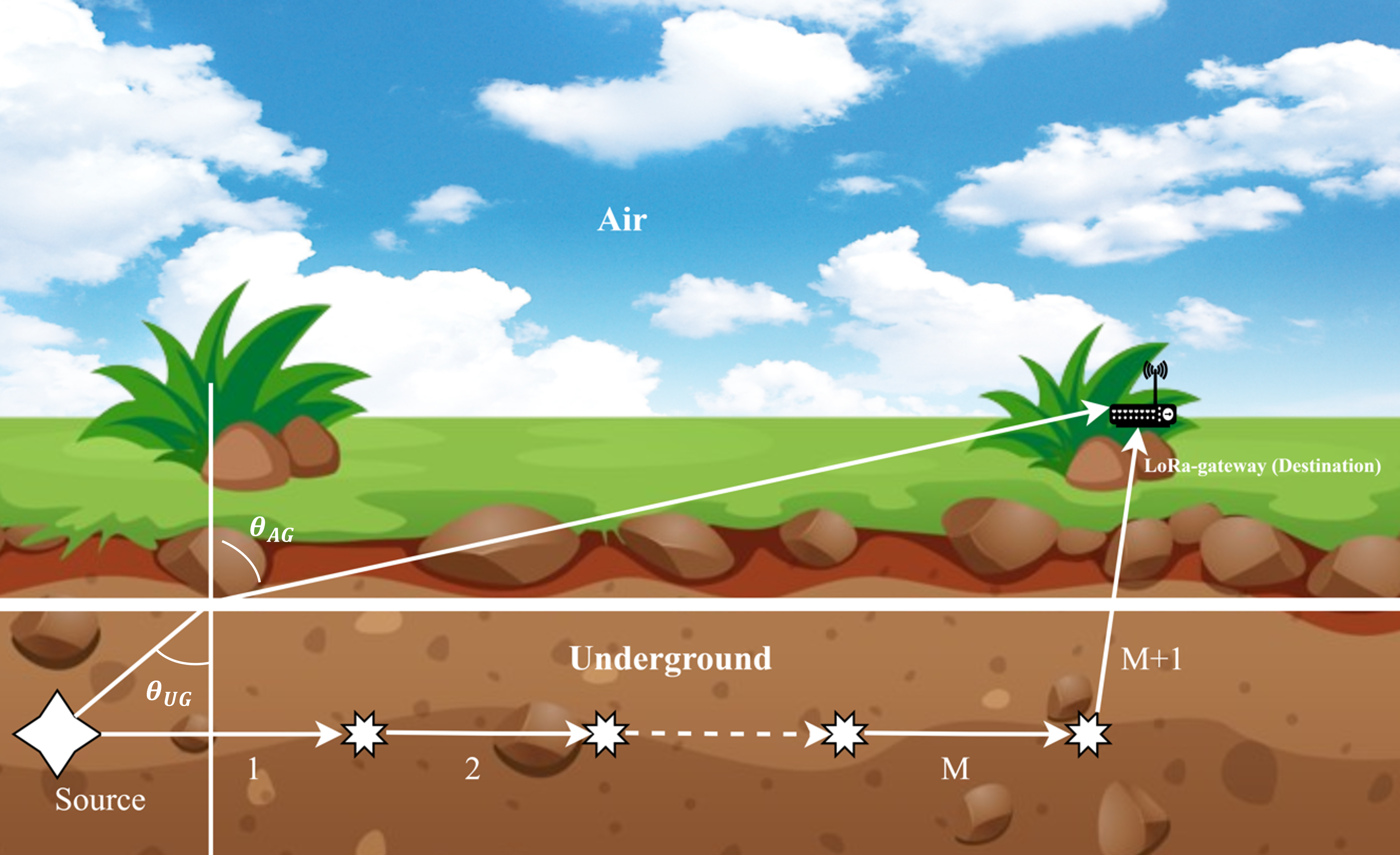}
\caption{An Illustration of the LoRa-based multihop IoUT network model.}\label{proposedmodel}
\end{center}
\end{figure}

However, research focusing on routing protocols and relaying in this domain is still being determined. In \cite{8603641}, the authors developed a routing method utilizing Ad-hoc On-demand Distance Vector (AODV) for LoRa devices, facilitating the establishment of a long-chain topology for data transfer in smart grids. Furthermore, they implemented a multi-hop technique to enhance packet reception rates by determining the most efficient path with the least hops along power transmission lines. An illustration of such a multihop LoRa-based IoUT network model is provided in Fig. \ref{proposedmodel}.
Addressing the communication challenges posed by LoRa technology in smart cities and farm environments, \cite{choi2020reliability} demonstrated that utilizing a multi-hop approach with LoRa technology proved more effective in large-scale smart farm environments. Implementing an Automatic Repeat Request (ARQ) system was essential to address the unexpected drop in the forward rate caused by the growing number of IoT devices. Furthermore, in \cite{dziri2016performance}, the authors investigated various relaying systems, mainly Decode-and-Forward (DF) relaying and Amplify-and-Forward (AF) relaying for terrestrial networks.
DF relaying involves the relay receiving and decoding the signal from the source node and re-transmitting it over a secondary channel. In contrast, AF relaying employs simpler relays that do not perform decoding. This relaying method is very effective in increasing the signal strength, but it has the drawback of amplifying noise and interference. However, \cite{kisseleff2015capacity} concluded that the DF relay mechanism is crucial in improving link dependability, primarily by reducing the spread of noise in underground environments.
The authors in  \cite{ma2019antenna} introduced a WUSNs that makes use of Magnetic Induction (MI) and relays to attain high performance using DF relaying, \cite{ma2019antenna} albeit they did not consider the performance analysis of multi-hop relays within LoRa-based IoUT networks.

\begin{figure}[h!]
\centerline{\includegraphics[width = 1\columnwidth]{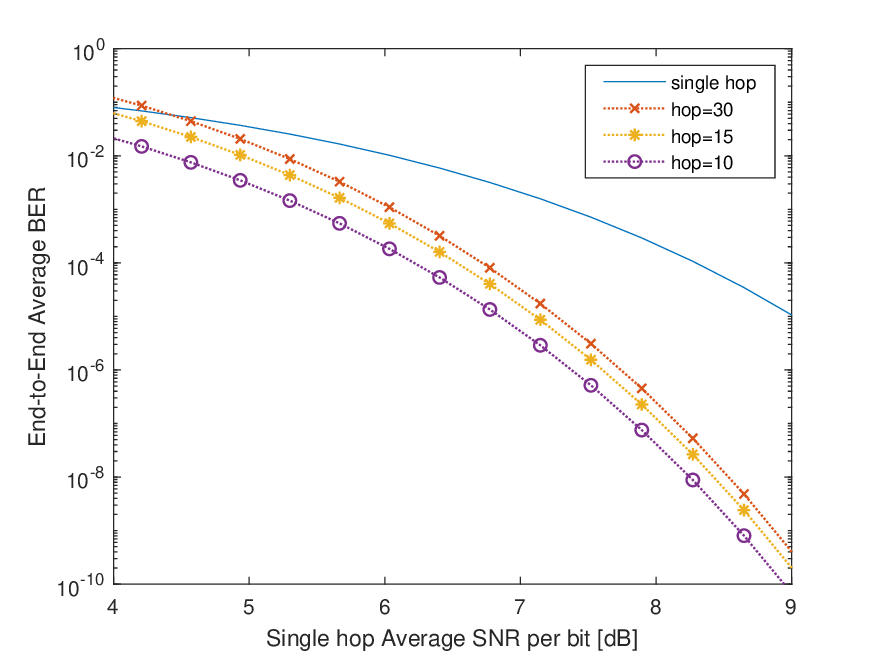}}
\caption{Average BER vs. SNR per bit for single- and multi-hop scenarios.}
\label{fig:multivssingle}
\end{figure}

It is crucial to attain extended communication ranges and enhance reliability to enhance the practicality of LoRa technology for real-world agricultural applications. Implementing multi-hop scenarios is essential for expanding coverage and maintaining reliable connectivity. Consider a subterranean IoT network leveraging LoRa wireless technology for communicating from an underground initial node to an aboveground target node through M number of relay nodes as depicted in Fig. \ref{proposedmodel}. To understand the effectiveness of LoRa in subterranean communication, it is crucial to obtain the expression for the probability of error. Based on \cite{morgado2010end}, the end-to-end average Bit Error Rate (BER) of the signal at the destination node is expressed as
\begin{equation}
    \mathbb{P}_{M+1} = \frac{1}{2} \times(1-(1-2\mathbb{P})^{M+1}),~~
  \forall M \geq 1,
    \label{eq:multi}
\end{equation}
where $M+1$ is the number of hops, $\mathbb{P}$ is the average BER for the single hop. The BER for a single hop is obtained by curve fitting method \cite{reynders2016chirp} and is given as
\begin{equation}
    \mathbb{P}_e = \frac{1}{2}\times Q(1.28 \sqrt{K{\gamma}}-1.28 \sqrt{K} + 0.4),
\end{equation}
where $K$ is the SF of Lora from 7 to 12, and $\gamma$ is the energy per bit-to-noise ratio. $\gamma$ is calculated as
\begin{equation}
    {\gamma}=\frac{\hat\gamma \times B}{R_b},
\end{equation}
where $\hat{\gamma}$ is SNR, $B$ is the BW, and $R_b=K\frac{B}{2^K}$ is the bit rate. The  SNR is calculated  as \cite{mcvarunchannelmodel}
\begin{equation}
    \hat\gamma = P_t - \ell_t - P_n,
\end{equation} where $P_t$ denotes the transmit power, $\ell_t$ is the total path loss, and $P_n$ represents the noise energy. The values of $P_t$ and $P_n$ can be  30 dBm and -103 dBm, respectively  \cite{mcvarunchannelmodel}. Fig. \ref{fig:multivssingle} demonstrates that the suggested multi-hop method surpasses the single-hop BER for a fixed value of SF of 7. Our analysis considered multi-hop links, with each hop demonstrating the same statistical behavior and an equivalent BER as a single hop.

In the case of the single hop, the transmission occurs directly between the source and destination nodes, covering a distance of $d$. This leads to a bad SNR and, consequently, a high BER. On the other hand, by including ($M +1$) hops, the distance between each node is reduced to ${d}/{M+1}$, resulting in a higher average SNR and improved BER. Therefore, in multi-hop IoUT networks, DF relaying greatly improves the BER  compared to a single-hop arrangement. Moreover, in multi-hop settings, there is a noticeable decrease in performance as the number of hops increases, even while the average SNR remains constant.

\section{Practical Implementation of LoRa-Based Agriculture IoT Networks}\label{practicalimplementation}
In this section, we discuss the extensive research surrounding the development and practical application of LoRa technology. The literature on LoRa development includes theoretical insights, experimental validations, deployment case studies, and performance assessments across various application fields. To maximize performance and efficiency in real-world circumstances, we evaluated and investigated the core features of LoRa technology, including modulation techniques, signal processing algorithms, and Radio Frequency (RF) propagation characteristics.

The study by \cite{9014994} explored utilizing WiFi and LoRa technologies for rescue monitoring. This concept established a connection between IoT devices and the internet through LoRa GWs and a WiFi router. The data generated by sensors on IoT modules can be accessed via a web or mobile application. The project incorporated a nano-gateway that links IoT modules to wearable devices using LoRa. The communication process encompassed determining network availability, collecting metrics, establishing a transmission channel, and transmitting data in the appropriate format. Both modeling and real-time trials indicated that, as per the findings in \cite{9014994}, LoRa emerged as the optimal choice for IoT-based monitoring applications. In a related study, \cite{9279019} investigated the deployment of an IoT network (SysAgria) in a vast agricultural region in Southwest Romania. SysAgria empowers farmers with proactive monitoring capabilities for soil, air, and light conditions influencing plant development. While the system proved ideal for small and medium-sized farms, larger agricultural establishments required a more comprehensive evaluation due to technological complexities and the demand for competitive solutions. The adoption of LoRa for communication is justified in  SysAgria based on its low power consumption, extensive coverage, cost-effective transmission, and resilience in noisy conditions.

The study by \cite{9103815} delves into a low-power farm environment detection system utilizing LoRa wireless technology for enhanced data collection and processing in intelligent agriculture. This system, designed for production sites, encompassed detecting soil parameters, modeling environmental conditions, and implementing control mechanisms such as lighting, irrigation, cooling, and man-machine interfaces through sensors. The system architecture comprises three layers, including the user access platform, the network layer, and sensors. Employing the LoRa protocol, the AS32TTL100 communication module with four communication modes facilitated effective monitoring and management of greenhouse farms. Laboratory testing and on-site data collection affirmed the practicality of the technology, showcasing increased agricultural output and reduced water usage.

In a related context, \cite{8834788} presented an IoT system utilizing LoRa for simplified monitoring of irrigated plants in agriculture. Sensors measuring the water potential of Hydrogen (pH), soil pH, and soil moisture communicate with an Arduino micro-controller over LoRa 915 MHz frequency. Their system aimed to prevent erratic water consumption and adverse weather conditions, demonstrating favorable impacts on plant features, particularly crops like pakchoi. Overall, the IoT-LoRa system enhanced agricultural operations by ensuring crops receive water with the proper pH \cite{8834788}.
The transformative impact of IoT in smart agriculture was highlighted in \cite{khoa2019smart}, which introduced a sensor node topology designed with cost-effective components for water management. By utilizing LoRa LPWAN technology, the sensor network was tested in a research lab, demonstrating its potential for both manual and automated farm management through a mobile application. This approach reduced costs while enhancing agricultural productivity, energy efficiency, and overall operational effectiveness.

In \cite{9831521}, a LoRa-based IoT system for smart farms was introduced as part of the progression into Industry 4.0, integrating Programmable Logic Controllers (PLC) with a wireless network of LoRa sensors. The system included a cloud-based monitoring application for remote visualization and control of the agricultural environment. Key components, such as sensor nodes, control equipment, a cloud server, and a web application platform, formed the system's core. The study emphasized the system's versatility, allowing users to build an IoT network, add devices, analyze data, and facilitate direct message exchange via messaging apps through a Telegram bot. In another study \cite{8394568}, the fusion of LoRa communication technology with intelligent Wireless Sensor Networks (WSN) is explored. The platform presented in this paper aimed to integrate IoT and knowledge communication technologies in agriculture and employed sensor accuracy analysis to create multi-functional sensor components. The study emphasized the role of LoRa technology in enabling outdoor data collection, such as agricultural or traffic signals in smart cities. It evaluated sensor accuracy for designing cost-effective, long-range sensing and communication systems for smart agriculture.

In \cite{9219028}, the study explored long-range communications for agricultural applications, focusing on the low-cost deployment of wireless environmental sensor nodes utilizing LoRa technology. These sensor nodes included soil moisture, temperature, humidity, rainfall, and light energy sensors. The study demonstrated the feasibility of LoRa for real-time, remote environmental monitoring in the field, achieving over 90\% accuracy in water measurements within a range of approximately 600 meters. The paper provided valuable insights into the potential of LoRa for efficient and accurate environmental monitoring in agriculture.

\newcolumntype{C}{>{\arraybackslash}X} 
\setlength{\extrarowheight}{1pt}
\begin{table*} [h!]
 \caption{Summary of the literature on the practical implementation of LoRa in agriculture. }
\label{table0110}
 \begin{tabularx}
{\textwidth}{|p{1.5cm}|p{4.5cm}|p{4.5cm}|p{5.9cm}|}
\hline
\hline
\textbf{References} &  \textbf{Technology} & \textbf{Applications} & \textbf{Possible Research Directions} \\
\hline
\cite{9279019} & LoRa & Agriculture monitoring, soil moisture, air, and light conditions monitoring in Southwest Romania  & Future research can lead to multi-sensor integration for effective agriculture monitoring  \\
\hline
\cite{9103815} & LoRa & Detection of soil quality and moisture, environmental parameters, lighting control, irrigation control, cooling control, and man-machine control & Future research can lead to optimizing IoT capabilities in terms of energy efficiency and integration of AI \\
\hline
\cite{8834788} & LoRa & measuring water pH, soil pH, and soil moisture, crop monitoring, adverse weather conditions in agriculture-IoT & Future research can include improving modulation efficiency and optimizing LoRa parameters (BW, SF) \\
\hline
\cite{khoa2019smart} & LoRa/LoRaWAN along with PLC & Monitor humidity conditions, Remote farm monitoring, Automate some tasks such as irrigation, or adjusting the
the temperature inside the warehouse  & Future research can lead to the utilization of ML algorithms for more effective agriculture monitoring \\
\hline
\cite{9831521} & IoT, LPWAN, LoRa, cloud computing, Web-based platforms
 & Smart farming, warehouse monitoring, remote device management, and automation of irrigation systems & Possibility to enhance automation with ML algorithms and optimizing water and energy usage in agriculture    \\
\hline
\cite{8394568}  &  LoRa technology with multi-sensor components & Agriculture monitoring, farms, and crop production monitoring &  Future research can investigate the integration of ML algorithms, an improved sensor accuracy, and energy harvesting  \\
\hline
\cite{9219028} & LoRa, Sigfox, LTE-M, NB-IoT & Smart cities, smart agriculture, wearables, asset tracking & Possibility towards more enhanced energy management, 5G integration, and LoRa network optimization  \\
\hline
\cite{9831210} & LoRaWAN with multi-sensors & Smart farming, crop management, environmental monitoring & Possibility towards multi-channel LoRa modules allowing more devices to connect simultaneously, and achieve reliable communication \\
\hline
\cite{gresl2021practical} & Sensors-based LoRa architecture along with data exchange protocols & Smart agriculture applications such as leaf wetness, humidity monitoring, plant disease modeling, improving crop quality, reducing production costs & Future research can lead to exploration of other ML algorithms such as RL \\
\hline
\cite{9108957} & SigFox  & Smart farming with sensing temperature, humidity, pressure, lightning, real-time data collection, and analysis &  Future research can lead to exploring more applications such as irrigation systems, monitoring chemical content in plants, and addressing security concerns  \\
\hline
\cite{doshi2019smart} & IoT, DHT11 temperature and humidity sensor, and soil moisture sensor
 & Smart farming, crop monitoring, temperature and moisture monitoring, smart irrigation systems, precision agriculture & Possibility to further enhancing IoT capabilities in agriculture, integration of AI, and implementing RL for different agriculture applications   \\
\hline
\hline
\end{tabularx}
\end{table*}

As the LoRaWAN standard was developed to meet the need for low-cost, low-power IoT applications with broader coverage than existing wireless technologies, the authors in \cite{9831210} sought to establish a low-power LoRaWAN-based network for smart farming. This network integrated Ultraviolet (UV) and Infrared (IR) sensors to measure air temperature, relative humidity, soil moisture, and sunshine intensity. 
The authors concluded that LoRaWAN, despite its slower data rate, was more energy-efficient and offered superior distance coverage compared to other technologies, making it well-suited for smart farming applications. 

In \cite{gresl2021practical}, the authors explored IoT implementation in agriculture through Smart Sensing for Agriculture (SSA), which combined low-cost, energy-efficient sensors with consumer electronics, utilizing communication protocols such as LoRaWAN, Bluetooth Low Energy (BLE), and WiFi Direct. SSA applications included greenhouse monitoring, smart irrigation, and animal tracking. The paper emphasized the importance of employing various communication protocols and emerging technologies, such as fog computing and Bayesian neural networks, to manage the large volumes of sensory data generated. An experimental network deployment tested the LoRa exchange protocol, connecting nodes to gateways (GWs). The authors concluded by highlighting the potential of SSA and the need to address challenges for broader implementation in agriculture. In \cite{9108957}, the authors investigated ways to prolong the battery life in a prototype that monitors open-field food storage with LoRaWAN technology. The study focused on challenges associated with transmitting data over considerable distances and limited battery capacity. Significant findings include scrutinizing sleeping current for battery lifetime elongation, low self-discharge batteries' significance, and sensor efficiency's instrumental role. The research provided crucial insights into IoT solutions within agriculture settings, particularly concerning optimal post-harvest conservation practices in open-field setups.

In \cite{doshi2019smart}, integrating IoT technologies into agriculture was explored to address challenges such as water scarcity and precise water resource monitoring. The study emphasized the role of WSN and IoT sensors in monitoring crop and soil conditions, introducing a low-power, long-range LoRa-based IoT system designed for real-time soil water monitoring. The goal was to develop a field-installable, user-friendly prototype model suitable for farmers with limited technological expertise.

Focusing on urban agriculture in smart cities, \cite{abioye2020review} introduced precision agriculture, leveraging IoT and ML technologies for efficient water management. The paper emphasized the integration of LoRaWAN and ML algorithms to optimize irrigation planning, reduce water wastage, and improve crop yields. It also explored the challenges of conventional irrigation practices and contributed to Agriculture 4.0 by proposing innovative water management solutions for urban agriculture through advanced technologies.

Lastly, \cite{mishra2023energy} outlined the design and implementation of a three-tier LoRa-based smart farming system, consisting of a sensing layer, a fog layer, and a cloud layer. This system addressed connectivity issues in rural areas, highlighting the significance of low-power LoRa components for connecting energy-efficient IoT devices to edge servers. The study offered insights into the advantages of LoRa technology over other communication technologies, particularly in regions with intermittent Internet access, demonstrating its suitability for smart agriculture applications.

This section examined the efficacy of LoRa-based agricultural IoT connectivity through a critical review of pertinent research findings. Numerous scholarly works showed the utility of LoRa technology in agricultural contexts, citing its contributions to real-time monitoring, resource optimization, field operations, and burgeoning trends within the domain. Central insights from these studies underscore LoRa's efficacy in mitigating challenges such as low power consumption and cost considerations. Furthermore, research illuminates its applicability across domains encompassing land management, irrigation, and environmental monitoring, thereby fostering the evolution of smart agricultural practices. In a nutshell, the literature underscores the transformative potential of LoRa technology, positioning it as a pivotal instrument in advancing data-driven and sustainable agricultural methodologies. Table \ref{table0110} summarizes the contributions of the works on practical implementations of LoRa in the agriculture sector.

\section{Comparison of LoRa with other Wireless Communication Technologies for Agriculture IoT}\label{comparison1}
LoRa stands out among various wireless communication technologies in smart agriculture due to its exceptional range and low power consumption. It is highly suitable for extensive agricultural fields requiring remote monitoring. Unlike WiFi and Bluetooth, which offer limited coverage, LoRa enables long-distance communication, allowing efficient data transmission across vast farm areas. Furthermore, compared to cellular networks, LoRa provides a cost-effective solution without expensive infrastructure, making it accessible for farmers seeking to implement IoT solutions for improved crop management and monitoring. Its adaptability to diverse agricultural environments and its capacity for transmitting small amounts of data over long distances underscores LoRa's potential to revolutionize smart farming practices by offering a scalable, energy-efficient, and economically viable communication platform. In the following, we compare various wireless communication technologies to LoRa for different applications in smart agriculture.

\subsection{WiFi}
Several research works have suggested integrating WiFi technology into various innovative agriculture applications \cite{akram2021smart,aliev2018internet,venu2022smart}. WiFi, operating on the IEEE 802.11 standard and utilizing RF bands, facilitates communication with a range of 20 to 100 meters \cite{sadowski2020wireless}. Its adoption enables efficient data transmission and connectivity within agricultural networks. Scholars have explored WiFi-enabled solutions to enhance monitoring, communication, and data exchange in agricultural contexts \cite{memon2016internet, hsu2020creative, muangprathub2019IoT}. For example, Vijayakumar et al. \cite{vijayakumar2015real} introduced an IoT module leveraging WiFi for real-time water quality monitoring, illustrating WiFi's versatility in agricultural sensor networks.
Efforts to optimize energy consumption and network longevity have led to innovations such as low-energy WiFi implementations in the data link layer \cite{lavanya2020automated, sarangi2014development}. These adaptations harness WiFi's energy-efficient characteristics, offering advantages such as reduced overhead, power-friendly communication, improved synchronization, and the utilization of short MAC frames. Moreover, adopting cost-effective WiFi microchips like the ESP8266 module \cite{gsangaya2020portable, song2020study} has facilitated internet connectivity in smart agriculture setups.
WiFi's versatility extends to forming long-distance networks, exemplified by the WiLD network that connects remote rural areas economically \cite{ahmed2018internet, widjaja2021IoT, le2021IoT}. Compared to LoRa technologies such as SigFox and NB-IoT, WiLD offered substantial data transmission and fog computing capabilities at the cost of higher energy consumption.

With a long-range antenna, the WiFi system investigated in \cite{10478705}, exhibited an impressive coverage range of up to 3 km, rendering it suitable for large-scale agricultural applications. However, signal degradation due to obstacles such as trees and vegetation may lead to packet errors. While WiFi entails a higher cost, approximately \$50 per device, it offers direct compatibility with a diverse array of data formats, including high-resolution images and videos. Conversely, LoRa presents a more economical option at \$5 per unit, featuring sufficient range to cover a 5-acre area with minimal energy consumption. In \cite{lin2023csi}, the authors explore the potential of radio-frequency Wireless Energy Transfer (WET) for underground sensor networks where they compare various Channel State Information (CSI)-free multi-antenna WET schemes and propose a distributed CSI-free system for efficient charging of underground sensors equipped with an appropriate number of antennas. Nonetheless, LoRa's efficacy in transmitting multimedia data is limited, necessitating intricate encoding and decoding techniques. In a nutshell, LoRa emerges as a cost-effective and energy-efficient solution for specific long-range communication contexts. In contrast, WiFi remains preferable for extensive coverage and robust multimedia data transmission with high cost and complexity. Compared to WiFi, LoRa offers distinct advantages for smart agriculture applications. It boasts an extended range for vast fields, low power consumption for battery-powered sensors, and strong signal penetration through obstacles and interference. These features deliver reliable connectivity for remote farms, unlike WiFi's limitations.

\subsection{Zigbee}
Adoption of ZigBee technology is significantly impacting smart agriculture, as evidenced by numerous research endeavors leveraging its wireless communication capabilities \cite{hidayat2020method,hidayat2017internet,mahir2018soil,kirtana2018smart}. Approximately one-sixth of studies within the IoT ecosystem have integrated ZigBee technology, capitalizing on its foundation on the IEEE 802.15.4 standard and operation in the 2.4 GHz Industrial, Scientific, and Medical (ISM) band. The categorization of ZigBee devices into EDs, routers, and coordinators highlights a structured approach to network organization and data transmission efficiency in agricultural contexts \cite{encinas2017design, nikolidakis2015energy}.
ZigBee provides a robust framework for connectivity between farming applications. Differentiating device types within ZigBee networks EDs, routers, and coordinators—is pivotal in facilitating data transmission and network management. In \cite{morais2008zigbee}, Morais et al have explored ZigBee's potential, utilizing devices like MPWiNodeZ to form efficient ZigBee networks, demonstrating the practicality and scalability of this technology in agricultural settings.

The versatility of ZigBee technology is evident in its applications for sensor networks in smart agriculture, offering scalability and simplified maintenance of interconnected sensor nodes. Additionally, studies such as \cite{kumar2014energy, fourati2014development} have demonstrated the efficacy of Xbee modules for data transmission across various environmental conditions. ZigBee's ability to establish reliable communication links over considerable distances proves invaluable for water quality monitoring and intra-sensor communication in greenhouses or irrigation systems. The role of ZigBee coordinators as central control stations underscores their significance in efficiently managing data flow and network operations, as emphasized by research efforts \cite{sinha2019architecting, patil2016early} in establishing robust ZigBee networks for agricultural applications.

Although LoRa and Zigbee are prevalent wireless communication technologies in smart agriculture applications, they exhibit notable differences in range, data rate, and power consumption. For instance, the study in \cite{10478705} highlighted that long-range, energy-efficient applications could benefit from LoRa's significantly extended communication range, covering up to a 5-acre area. Operating within the 868 MHz frequency range and featuring a data rate of 37.5 Kbps, LoRa contrasts with Zigbee, utilizing the IEEE 802.15.4 PHY standard, which offers more restricted wireless coverage with an impressive 1.2 km outdoor RF line-of-sight range and a higher data rate of 250 Kbps. Despite its lower unit cost of \$5 compared to Zigbee's \$16, LoRa faces challenges in transmitting multimedia data efficiently due to severe delays and packet loss, particularly with high-resolution photos. Conversely, Zigbee needs help directly transmitting images, necessitating encoding and decoding methods, thus complicating the data acquisition. The choice between LoRa and Zigbee depends on the specific requirements of the application. LoRa is ideal for reliable, long-range communication. On the other hand, Zigbee excels in localized, dependable data transfer.

\subsection{RFID}
RFID technology emerged as a transformative force in smart agriculture IoT, revolutionizing data capture and object identification processes \cite{rayhana2021rfid, ruiz2011role, agrawal2019smart, wasson2017integration}. RFID enhances efficiency and precision in agricultural operations by assigning unique identifiers to objects and enabling seamless communication between tags and readers. Tags with distinct ID numbers and environmental data communicate with readers through radio waves, enabling precise object identification \cite{shi2019mobile}. RFID technology's versatility in agriculture spans various applications, from enhancing livestock management systems to optimizing supply chain operations. Noteworthy studies, such as the mobile measuring system for pig body components by Shi et al. \cite{shi2019state} and the utilization of RFID in dairy farm management by Trevarthen et al. \cite{trevarthen2008rfid}, exemplify the diverse capabilities of RFID in agricultural settings.
Innovative applications of RFID technology extend to tracking and monitoring crucial aspects of agricultural processes, as demonstrated by Aguzzi et al. \cite{aguzzi2011new} and Kim et al. \cite{kim2014rfid} in developing RFID tracking solutions for studying the emergence patterns of marine crustaceans. Additionally, the implementation of RFID for ensuring traceability in food supply chains, explored by Kelepouris et al. \cite{kelepouris2007rfid}, and Ruan et al. \cite{ruan2016monitoring}, highlights its role in enhancing transparency and efficiency in agricultural production and distribution. The seamless integration of RFID technology into smart agriculture IoT systems streamlines data collection and management, paving the way for enhanced operational efficiency and decision-making in modern farming.

LoRa and RFID are two popular connection systems in smart agriculture, each with unique benefits. While RFID is easy to install, LoRa is better at long-range transmission and energy efficiency, which makes it perfect for large agricultural landscapes. Despite its initial setup complexity, LoRa is relatively cost-effective in large-scale applications because of its continuous data transfer over great distances and decreased power consumption. On the other hand, RFID is easier to use and more appropriate for applications involving monitoring and identification at a close distance in small spaces. Despite somewhat greater upfront costs than RFID, LoRa's efficiency and range make it more cost-effective over the long term. The decision between LoRa and RFID ultimately comes down to many criteria, such as distribution requirements, efficiency requirements, resilience of the original design, and long-term cost evaluations. Both protocols represent the dynamic nature of decision-making in smart agriculture by providing customized solutions appropriate for various agricultural conditions and operational demands.

\subsection{Cellular Systems}
In smart agriculture, IoT, various generations of mobile communication standards—2G, 3G, 4G, 5G—play crucial roles in facilitating connectivity between IoT devices and cellular networks \cite{avcsar2022wireless, ray2017internet, tomaszewski2023mobile, tang2021survey}. Researchers such as Saqib et al. \cite{saqib2020low} have utilized 2G modules for weather data gathering and cloud connections using Global System for Mobile Communications/General Packet Radio Service (GSM/GPRS) protocols, creating cost-effective information monitoring systems tailored for smart farming applications. The selection of mobile communication standards is a strategic decision. For example, Namani and Gonen \cite{namani2020smart} utilize 4G technology to enable high-speed communication between controllers and servers, whereas 2G modules are typically favored in rural agricultural regions where higher-generation networks are not yet fully deployed \cite{damsgaard2022wireless}.
Factors such as coverage, data rates, and suitability for remote agricultural locations drive the selection of mobile communication technologies. Among cellular technologies, GPRS emerged as a versatile option for Machine-to-Machine (M2M) communication in agriculture \cite{bejgam2021integrating, sudarmani2022machine}. Its widespread availability and cost-effectiveness make it practical for sensor-to-base station communication, particularly in scenarios like greenhouse monitoring, where remote data transmission is essential for efficient agricultural management \cite{mekala2017novel}.
Recent advancements in 5G technology have revolutionized farming practices in smart agriculture IoT \cite{damsgaard2022wireless, tang2021survey, valecce2019interplay, murugamani2022machine}. With its high-speed, low-latency capabilities, 5G facilitates advanced applications such as remote monitoring, autonomous machinery, and precision agriculture, ultimately boosting productivity and sustainability.

When assessing cellular communication and LoRa for smart agriculture, Cellular Communication stands out for its extensive coverage. It is particularly suitable for covering large agricultural regions, especially in areas with well-established community infrastructure. However, this convenience often entails ongoing expenses associated with data plans, making it a consideration for budget-conscious farmers. Furthermore, implementing cellular communication may incur higher energy consumption, potentially affecting the battery life of devices, and involves complexities such as managing network carriers and configuring devices.
In contrast, LoRa emerged as a cost-effective solution with remarkable long-range capabilities. LoRa's energy efficiency is particularly notable, enabling devices to operate on battery power for extended periods, which is essential for remote monitoring in agricultural settings. Moreover, deploying LoRa networks is generally simpler, with standardized protocols and minimal configuration requirements, making it accessible even to users without advanced technical skills.

\newcolumntype{C}{>{\arraybackslash}X} 
\setlength{\extrarowheight}{1pt}
\begin{table*} [h!]
 \caption{Comparison of Lora with other wireless Communication Technologies for Agriculture IoT}
\label{table2}
 \begin{tabularx}
{\textwidth}{|p{2cm}|p{3cm}|p{1.95cm}|p{7cm}|p{2cm}|}
\hline
\hline
\textbf{Technology} &  \textbf{Type of Communication} & \textbf{Range} & \textbf{Applications in Smart Agriculture} & \textbf{References} \\
\hline
WiFi & Wireless IEEE 802.11 standard (2.4 GHz  and 5 GHz ) & 20 to 100 m & Monitoring climate conditions, precision farming, crop management, livestock tracking, and irrigation control  & \cite{akram2021smart,aliev2018internet,venu2022smart,sadowski2020wireless,memon2016internet, hsu2020creative, muangprathub2019IoT,vijayakumar2015real,lavanya2020automated, sarangi2014development,gsangaya2020portable, song2020study,ahmed2018internet, widjaja2021IoT, le2021IoT}   \\
\hline
Zigbee & Wireless IEEE 802.15.4 standard (2.4 GHz ISM band  and 5 GHz ) & 10 to 100 m & Smart monitoring, agrochemical applications, disease detection, crop monitoring and management, smart water management, and greenhouse automation  & \cite{hidayat2020method,hidayat2017internet,mahir2018soil,kirtana2018smart, encinas2017design, nikolidakis2015energy,morais2008zigbee,kumar2014energy, fourati2014development,sinha2019architecting, patil2016early}   \\
\hline
RFID & Wireless IEEE 802.15.4 standard (LF: 125-134 kHz, HF: 13.56 MHz, UHF: 433 MHz, 860-960 MHz, and 2.45 GHz) & Passive RFID up to 10 m, Active RFID up to 100 m & Livestock management, crop monitoring, supply chain management, automated irrigation systems, tool tracking, and greenhouse automation  & \cite{rayhana2021rfid, ruiz2011role, agrawal2019smart, wasson2017integration,shi2019mobile,shi2019state,trevarthen2008rfid,aguzzi2011new, kim2014rfid,kelepouris2007rfid,ruan2016monitoring}   \\
\hline
Cellular systems & Wireless IEEE 802.11 (Up to 2GHz for 3G, up to 8 GHz for 4G, and up to 30GHz for 5G) & large distance up to 10 km  & Crop monitoring,  pests or disease detection, drone spraying, irrigation, and fertilization monitoring   & \cite{avcsar2022wireless,sushanth2018IoT,ray2017internet,tomaszewski2023mobile,tang2021survey,saqib2020low,namani2020smart,damsgaard2022wireless,bejgam2021integrating,sudarmani2022machine,ilyas2020smart,mekala2017novel,valecce2019interplay,murugamani2022machine}   \\
\hline
Bluetooth & Wireless IEEE 802.15.3-2003 and IEEE 802.15.3b-2005 (MAC and PHY) & Bluetooth v2.1 is up to 100 m, Bluetooth 4.0 (LE) up to 100 m, and Bluetooth 5 (LE) is up to 400 m & Animal health tracking, crop monitoring, inventory management, reducing waste, ensuring food safety, optimizing humidity and temperature, health monitoring of cattle, fertilizers management, pesticides control, and crop harvesting detection &  \cite{mini2023IoT, heble2018low, al2022smart, kim2008remote, ilapakurti2015building,hasan2021smart,monica2017IoT, farooq2019survey,  ayaz2019internet}  \\
\hline
LoRa & IEEE SA - P1451.5.5 (150 MHz to 1020 MHz)  & Up to 5 km (urban), up to 15 km (suburban), and up to 45 km (rural)  & Sustainable  farming, soil irrigation monitoring, crop monitoring,  cattle tracking, indoor botanical garden control, irrigation management, and remote monitoring  & \cite{8580479, miles2020study, arshad2022implementation,changqing2018internet,9220017,9533237,107324,siddique2019review,8269115,chen2020aIoT,tacskin2020long, codeluppi2020lorafarm, adami2020monitoring}   \\
\hline
\hline
\end{tabularx}
\end{table*}

\subsection{Bluetooth}
Bluetooth technology emerges as a low-power, short-range personal area network ideal for facilitating mobile communication within agricultural systems \cite{mini2023IoT, heble2018low, al2022smart}. For instance, Kim et al. \cite{kim2008remote} harnessed Bluetooth for sensor-based variable rate irrigation systems, emphasizing seamless integration of existing data loggers and sensors through cost-effective wireless communication modules. Bluetooth modules enable wireless data transmission between in-field sensing and central base stations, enhancing agriculture data collection and monitoring capabilities \cite{ilapakurti2015building, hasan2021smart}.
Bluetooth technology, particularly BLE frameworks, offers advantages to smart agriculture IoT applications. Overcoming limitations such as power constraints and radio range optimization is crucial for maximizing Bluetooth's effectiveness in agricultural systems \cite{monica2017IoT, farooq2019survey}. Integrating solar panels for battery recharging and optimizing power class and antenna configurations can mitigate power shortages and improve radio coverage, ensuring robust and reliable Bluetooth connectivity in agricultural environments \cite{ayaz2019internet}. By optimizing system designs and leveraging Bluetooth technology effectively, smart agriculture IoT solutions can overcome challenges and harness wireless communication's full potential for enhanced monitoring, data collection, and operational efficiency in farming practices.

Bluetooth and LoRa are the two leading connectivity platforms in smart agriculture, each offering unique benefits and considerations. In terms of deployment, Bluetooth is generally more straightforward than LoRa, making it more convenient for smaller applications. However, LoRa shines in terms of energy efficiency, boasting lower power consumption for battery-powered devices in remote agricultural environments. More importantly, LoRa's long-lived capabilities make it ideal for extensive agricultural landscapes, allowing for seamless data transmission over long distances. Given their complexity, LoRa systems can be more difficult to set up initially due to the need for GW devices and networks. In contrast, Bluetooth systems are generally simple and plug-and-play in nature. Finally, although Bluetooth devices are generally expensive up front, the cost-effectiveness of LoRa is reflected in larger applications, where its efficiency and range later can reduce operating costs in the long run. In short, the choice between Bluetooth and LoRa for smart agriculture is dynamic and depends on factors such as efficiency, distribution requirements, robustness of initial design, and duration to assess costs, with each protocol providing tailored solutions suited to different agricultural needs and environments \cite{10478705}.

\subsection{LoRa versus other Communication Technologies}
LoRa technology surpasses other wireless communication technologies in smart agriculture applications, such as Bluetooth, ZigBee, and Wi-Fi. Its long-range capability enables communication over vast agricultural fields and remote areas where traditional technologies may struggle to maintain connectivity. This extended range proves beneficial for applications like greenhouse monitoring, pest detection, and farm monitoring, where reliable communication over large distances is vital for effective data collection and management.
Moreover, LoRa technology boasts low energy consumption, making it a cost-effective and sustainable solution for smart agriculture IoT deployments. Operating on minimal energy requirements while providing connectivity over expansive agricultural landscapes sets LoRa apart from other technologies, ensuring efficient data transmission without draining resources.
Additionally, the flexibility of LoRa technology, with features like SF for customizable trade-offs between coverage area, data rates, and packet size, allows for tailored communication solutions meeting the specific needs of smart agriculture systems. This adaptability and efficiency make LoRa technology the preferred choice for agricultural applications requiring long-range connectivity, low power consumption, and reliable data transmission in challenging environments. As summarized in \cite{ koubaa2020smart}, LoRa technology has better efficiency and scalability in agricultural settings than Bluetooth, ZigBee, cellular systems, and Wi-Fi. A detailed comparison of LoRa with other communication technologies for smart agriculture IoT is provided in Table \ref{table2}.

Besides, LoRa-based underground communication emerges as a promising solution for long-range, low-power communication in demanding underground environments \cite{zhang2023loraaid,lin2020experimental,li2019large,guo2021internet,liu2023q}. Leveraging its high reception sensitivity and low power consumption, LoRa technology holds the potential to significantly enhance the quality of electromagnetic signal transmission and extend the operational lifespan of buried nodes. Through thorough research and analysis, various facets of LoRa-based underground communication have been investigated, encompassing channel models, propagation performance, and adaptive parameter selection methods, laying the groundwork for a more reliable and efficient underground communication system \cite{ebi2019synchronous,zhou2021impacts}.
Initial efforts focused on evaluating LoRa propagation in soil, followed by detailed assessments of its transmission performance in underground channels, theoretically and empirically. Moreover, a statistical underground channel model was developed to establish the relationship between LoRa transmission range and BER \cite{wan2017lora}. The potential of LoRa and LoRaWAN protocols for underground monitoring activities was explored, alongside investigations into channel models for UG2AG and AG2UG communication \cite{lin2019preliminary,di2021lorawan,moiroux2022evaluation}. These studies scrutinized the impact of propagation direction, PHY parameters, and burial depth on LoRa propagation performance through practical experimentation.
A novel adaptive method for selecting LoRa parameters to evaluate UG2AG link quality was proposed in \cite{lin2020experimental}, aiming to prolong wireless sensor nodes' longevity significantly. Additionally, a model presented in \cite{lin2020link} for a non-backfilled channel was developed to quantify the effects of propagation direction, burial depth, inter-node distance, and backfill conditions on link quality. Despite the advancements, the critical aspect of underground node communication over large distances through relaying still needs to be addressed in existing research.

\section{AI-enabled LoRaWAN for Agricultural IoT} \label{AI-LoRAWAN}
This section highlights key research efforts in advancing AI-enabled LoRaWAN for Agricultural IoT, demonstrating how the integration of AI and ML with LoRaWAN technology is revolutionizing precision agriculture. While there is a substantial body of literature on AI-based agricultural IoT networks, only a limited number of studies focus on integrating AI with LoRaWAN-based agricultural IoT. In the following we provide a detailed discussion of these works.

In \cite{qazi2022iot}, the authors provide a comprehensive review of the latest advancements and challenges in smart agriculture, focusing on the integration of IoT and AI technologies. Their work bridges the gap between theory and real-world implementation, offering insights into the practical applications of these technologies. The review also explores future trends, aiding farmers and researchers in making informed decisions on adopting advanced technologies to optimize agricultural practices.
In another study \cite{majumdar2024enhancing}, the authors investigate the role of 5G networks in enabling seamless, energy-efficient connectivity in Agriculture 4.0. The review examines 5G technologies, including HetNets, LPWANs, AI-driven data processing, and cloud computing, emphasizing their potential in automating agricultural operations. It also highlights the need for energy-efficient solutions, particularly for battery-powered IoT devices like UAVs and sensors, and suggests energy harvesting as a key area for future research.

The work in \cite{alkhayyal2024recent} explores the integration of ML and AI to optimize LPWANs by using AI-driven techniques, such as deep learning and reinforcement learning, to improve network performance, resource allocation, and energy efficiency. The study concludes that AI and ML significantly benefit resource management and energy consumption, contributing to the development of more efficient and sustainable IoT networks.
In \cite{farhad2023ai}, the authors address inefficiencies in resource allocation in LoRaWANs for both static and mobile IoT applications, particularly focusing on packet loss caused by incorrect spreading factors assigned to mobile devices. The proposed AI-driven resource allocation scheme, which employs a deep neural network, enhances the performance and reliability of LoRaWAN in dynamic environments, ensuring efficient and reliable communication in static and mobile settings.

The study in \cite{singh2022joint} leverages LoRaWAN-based IoT applications and AI techniques to enhance precision agriculture in greenhouses. Based on datasets collected over nine months, the research evaluates LoRaWAN performance and develops an AI-based monitoring technique for precision agriculture. The findings show that LoRaWAN signals, combined with AI, can predict plant growth with a 10\% error margin, demonstrating the potential for joint communication and sensing without additional hardware.
Lastly, \cite{farhad2023lorawan} explores key challenges in LoRaWAN, such as parameter configuration, interference, and optimized ADR. The authors emphasize that machine learning offers a promising solution for intelligent and efficient resource management, which is crucial for the broader adoption of LoRaWAN.

To summarize, AI-enabled LoRaWAN holds great promise for enhancing Agricultural IoT by optimizing network performance, resource allocation, and predictive capabilities. The studies reviewed here underscore AI's role in addressing challenges like energy efficiency, interference, and scalability. Continued research is essential to advance these AI-driven solutions, leading to more intelligent, efficient, and sustainable agricultural systems.

\section{Future Research Directions}\label{futuredirections}
Current research indicates the promising potential of LoRa-based IoT technologies to enhance agriculture through wireless connectivity in rural areas and deploying in-field smart devices. This paper delves into the opportunities a LoRa-based communication system presents for diverse agricultural applications. Nevertheless, several unresolved challenges demand attention. For instance, while LoRa exhibits efficacy over expansive agricultural landscapes, the intermittent availability of power sources necessitates energy self-sufficiency in nodes. Additionally, the proliferation of sensors in smart farming systems underscores the need for seamless integration of multiple communication protocols, particularly in environments featuring diverse suppliers. Addressing these outstanding challenges is imperative to enable the widespread deployment of LoRa systems in smart agriculture initiatives. The following highlights potential future research directions for LoRa-based agriculture IoT networks.

\subsection{Advanced Channel Models}
Exploring advanced channel models based on LoRa technology presents a promising avenue for future research in smart agriculture IoT. By investigating LoRa-based communication channels within agricultural environments, researchers can enhance data transmission reliability and efficiency for applications such as precision farming, environmental monitoring, and livestock management. Understanding the unique characteristics of LoRa channels in agricultural settings offers insights into optimizing network performance, extending communication range, and mitigating interference challenges to enhance the capabilities of smart agriculture IoT systems.
Recent technological advancements, exemplified by systems like IoT-AgriSens \cite{arshad2022implementation,vijayaraghavan2023IoT}, optimize signal propagation in challenging agricultural environments with varying terrain, vegetation, and weather conditions. Research efforts have focused on developing systems such as the LoRa-based smart agriculture decision support system to enhance crop yield through real-time monitoring. Recent proposals like IoT-AgriSens aim to effectively monitor field data and predict crop yields \cite{codeluppi2020lorafarm}. The intelligent sensors module in the IoT-AgriSens system incorporates various sensors, including temperature, humidity, NPK, soil moisture, soil conductivity, and pH sensors, utilizing LoRa technology for real-time monitoring of air temperature, humidity, soil temperature, moisture, and pH.


Looking forward, future research into LoRa-based advanced channel models within Agriculture 4.0 must address the specific technical challenges presented by agricultural environments. This includes a detailed analysis of signal propagation characteristics over heterogeneous terrains, crop densities, and varying atmospheric conditions and quantifying the impact of interference from both natural and man-made sources. Furthermore, understanding the limitations of LoRa's network scalability in large-scale agricultural deployments, particularly in handling dense sensor networks with stringent latency and data integrity requirements, needs considerable attention.

Future efforts should focus on developing environment-specific channel models to optimize dynamic spectrum management strategies, mitigating co-channel interference, and ensuring reliable, long-range data transmission in highly variable conditions. Accordingly, by addressing these technical aspects, the integration of LoRa with AI-driven decision support systems and predictive analytics platforms can be further optimized, facilitating precision agriculture applications such as autonomous machinery control, real-time crop monitoring, and predictive equipment maintenance. Thus, refining channel modeling for agricultural environments can enhance communication efficiency, improve data transmission reliability, ensure robust connectivity, and optimize resource management across diverse agrarian settings.

\subsection{Data Compression and Optimization}

Future research should explore innovative data compression methods for precision agriculture by utilizing hybrid, compressed sampling techniques and amalgamating exact and greedy approaches to devise lightweight and efficient data compression algorithms tailored explicitly for agricultural sensor data. Additionally, future studies need to emphasize data compression and optimization techniques to reduce transmission sizes and optimize network bandwidth utilization, enabling more efficient use of resources and extending the lifespan of IoT devices in agricultural environments. Future research into node clustering algorithms and data compression techniques should focus on enhancing the efficiency of data transmission and collection in precision agriculture applications. Investigating energy-optimized data flow modeling and its impact on network bandwidth utilization could lead to more effective data compression strategies, ultimately conserving energy in resource-constrained IoT devices. Simulation-based evaluations can further demonstrate the effectiveness of these methods in achieving bandwidth-efficient and energy-saving data collection, particularly in UAV-assisted agriculture monitoring, showcasing significant advancements in optimizing sensor networks for Agriculture 4.0.\cite{9210225}.

Furthermore, smart farming integrates cutting-edge technologies such as IoT and cloud computing into agricultural practices. However, implementing IoT-based tracking systems in agriculture presents challenges, mainly due to the lack of clear guidance for practitioners \cite{triantafyllou2019precision}. Therefore, future research should focus on developing clear guidelines and standardized frameworks for implementing these systems. Specifically, priority should be given to creating a standardized reference architecture model that incorporates advanced data compression and optimization techniques to address energy consumption challenges across its seven key layers. By integrating data compression methods, optimized bandwidth utilization, and environment-specific communication protocols, future solutions can ensure reliable, scalable, and energy-efficient IoT networks tailored to agricultural needs \cite{triantafyllou2019precision}.

\subsection{Standardization of Data Formats}

Future research in Agriculture 4.0 must focus on addressing the complexities of achieving seamless integration and interoperability across diverse devices, platforms, and data sources. As agricultural operations increasingly rely on various sensors, applications, and equipment from multiple vendors, there is a growing need to develop advanced interoperability techniques. These techniques should enable the integration of different devices, applications, and platforms, ensuring smooth data flow and effective collaboration between systems \cite{lanza2016proof}. One crucial direction for future research is the development of standardized protocols, Open Application Programming Interfaces (APIs), and unified data formats to facilitate seamless data exchange and enhance interoperability among diverse agricultural systems. These advancements can help farmers access a unified view of their operations, enabling more informed decision-making based on comprehensive and real-time data insights. In particular, establishing standardized data formats for LoRa devices and platforms can facilitate smooth data interchange across different agricultural applications. Interoperability standards, which have been successfully implemented in other industries, should be adapted to the agricultural sector to improve the efficiency of data sharing and integration. These standards can enable the seamless connectivity of LoRa devices from various manufacturers, a critical requirement for developing large-scale, comprehensive agricultural monitoring and control systems that incorporate diverse sensors and equipment \cite{bahlo2019role}.

\subsection{Localization and Positioning Techniques}
LoRa-based localization and positioning techniques offer a promising avenue for future research in smart agriculture IoT. These techniques aim to precisely track agricultural assets, such as equipment and livestock, without relying on costly Global Positioning System (GPS) solutions \cite{vijayaraghavan2023IoT}. Efforts in research can focus on integrating LoRa with advanced hybrid ranging techniques such as Time-of-Flight (ToF)/RSSI to enhance location accuracy and reliability in intelligent agriculture settings \cite{islam2024lora}. These advancements aim to overcome challenges related to real-time data acquisition for field monitoring and crop yield predictions, offering a more efficient and cost-effective solution for smart agriculture systems.
By exploring innovative methods to accurately determine the location of agricultural assets and resources using LoRa-based systems, researchers can enhance precision farming practices, optimize resource utilization, and improve overall operational efficiency in farm settings. Future studies can integrate LoRa technology with GPS, inertial sensors, and ML algorithms to create robust and reliable positioning systems for intelligent farming applications \cite{khalil2023energy}. Harnessing the long-range capabilities of LoRa for location tracking and asset management enables real-time monitoring of crop growth, livestock movements, and equipment utilization, leading to more informed decision-making and enhanced productivity in agriculture. Implementing LoRa-based localization solutions can revolutionize precision agriculture by providing farmers with accurate spatial data, facilitating targeted interventions, and optimizing resource allocation for sustainable and efficient farming practices.

\subsection{On-device AI models}


Future research in Agriculture 4.0 should focus on advancing the integration of AI and IoT to revolutionize smart farming by leveraging deep learning techniques and Artificial Intelligence of Things (AIoT) systems. These technologies can autonomously process large and complex datasets, enhancing their ability to make intelligent decisions related to crop management, resource optimization, and sustainability without human intervention \cite{9453402}. This convergence of AI and IoT can significantly increase the efficiency of agricultural operations, supporting precision farming and sustainable agricultural practices.

Moreover, traditional cloud-based processing encounters significant bottlenecks in rural areas due to high network latency and limited bandwidth availability, making edge computing a critical solution. Thus, future research should focus on developing advanced edge computing architectures that utilize distributed intelligence, enabling localized data processing at the sensor or gateway level. This can result in reduced reliance on centralized cloud infrastructure, minimized data transmission loads, and real-time analytics with ultra-low latency. In addition, integrating edge AI models can further enhance the system's ability to make autonomous decisions for tasks like precision irrigation, pest control, and crop health monitoring.

\subsection{Multi-Sensor Integration}
Plantation and crop monitoring necessitate integrating diverse sensors to facilitate effective management and enhance productivity, particularly in agriculturally heterogeneous regions \cite{9993728}. For example, some sensors enable an indirect assessment of plants' nutritional requirements at the point of application \cite{grisso2011precision}. Visual monitoring empowers growers to ascertain field conditions, avert insect infestations, and safeguard crops against potential harm inflicted by trespassers \cite{andrews2021remote}, thereby mitigating agricultural output losses. Despite the utility of such monitoring systems, challenges arise when deploying these devices wirelessly over expansive agricultural domains.  Integrating diverse sensor types, such as soil moisture sensors, weather stations, drones equipped with multispectral cameras, and satellite imagery analysis tools, can provide comprehensive insights into various aspects of crop health, soil conditions, and environmental parameters. This multifaceted approach enables growers to make informed decisions regarding irrigation scheduling, pest management, and crop health optimization. Additionally, integrating sensor data with ML algorithms can facilitate predictive analytics, allowing for early detection of potential issues and proactive intervention strategies. Consequently, the holistic integration of multi-sensor data streams coupled with advanced analytical techniques presents a promising avenue for enhancing agricultural management practices and driving sustainable productivity gains.

\subsection{Sustainable Energy Solutions}
Integrating LoRa devices powered by renewable energy sources such as solar, wind, and kinetic energy into agricultural IoT systems offers a sustainable alternative to reliance on traditional power grids and batteries. The use of diverse renewable energy sources, including solar, wind, thermal, water, and RF energy, is gaining traction in smart farming applications \cite{javaid2023self,shaikh2016energy}. With advancements in power electronics and transmission technologies, solar and wind have emerged as predominant energy sources for powering IoT devices in agriculture \cite{10112789}. However, effectively monitoring and managing these energy systems is essential for maintaining economic viability.

Future research in Agriculture 4.0 should focus on developing predictive maintenance algorithms, energy optimization strategies, and advanced energy management systems to enhance the seamless integration of renewable energy into LoRa-based IoT networks. Exploring innovative energy harvesting techniques (e.g., bio-energy from agricultural waste and kinetic energy from farm machinery) can diversify the renewable energy options available for agricultural applications. Additionally, integrating energy storage technologies such as super-capacitors and advanced batteries can mitigate issues related to intermittent power supply, ensuring consistent operation of IoT nodes in remote locations. Moreover, Magnetic Induction (MI) communication technology, known for its high data throughput, low latency, and efficient energy harvesting capabilities, can serve as an underwater node wake-up system. This technology can be adapted for LoRa-based wake-up modules, triggering on-demand sensing strategies that further optimize the energy efficiency of underground IoT nodes. Incorporating these wake-up mechanisms in agricultural IoT applications can extend the operational lifespan of IoT devices, reduce overall energy consumption, and contribute to more sustainable and productive farming practices.

\subsection{Energy Efficiency Protocols}
Exploring energy-efficient routing protocols for smart agriculture IoT systems utilizing LoRa technology presents a promising avenue for future research. The evolution of the LoRa PHY anti-frame loss mechanism is expected to reduce energy consumption and packet loss while enabling seamless communication over extended distances, enhancing data collection and analysis in agricultural settings \cite{hota2020energy,krivzanovic2023advanced}.
Future advancements in energy-constrained optimization for SF allocation in LoRaWAN are poised to revolutionize energy management by dynamically allocating SFs based on evolving constraints, optimizing network performance and resource utilization. Integrating combinatorial Multi-Armed Bandit (MAB) algorithm-based joint channels and SF selection into LoRa networks holds promise for maximizing energy efficiency through intelligent algorithms, paving the way for more sustainable and cost-effective innovative agriculture solutions.
As the agricultural IoT landscape evolves, implementing robust security measures against LoRaWAN PHY-based attacks will be crucial to safeguard data integrity and ensure energy-efficient operations. Furthermore, the future adoption of adaptive network techniques, such as the strategic use of IEEE 802.11ac for high data rate transfer and LoRaWAN for low data rate transfer, will play a pivotal role in tailoring energy consumption to specific network demands, thereby enhancing the overall efficiency and sustainability of smart agriculture systems.
Embracing these cutting-edge protocols and technologies holds immense potential for driving innovation, improving agricultural practices, and minimizing energy consumption in a rapidly evolving digital landscape.

\subsection{Long Range Reliable Communication}
Although LoRa offers exceptional capabilities for extended-range communication with low power consumption, making it ideal for covering large agricultural areas with minimal infrastructure, challenges such as scalability, interference, and reliability over long distances still require innovative solutions. Therefore, future research should prioritize optimizing LoRa networks for large-scale deployment across vast agricultural fields. Future research on long-range, reliable communication using LoRa must address real-world challenges such as fading channels, interference, and environmental obstacles, particularly in Non-Line-of-Sight (NLoS) conditions. Studies have demonstrated that LoRa's reliability in noisy, mobile, and obstructed environments common in agricultural settings can reach 90.23\%, highlighting the need for optimized configuration \cite{abdallah2024improving}. Therefore, future research should focus on improving LoRa's hardware, software, and protocols to enhance its dependability for long-range, reliable communication in these complex environments.

Moreover, future research must emphasize the exploration of advanced network topologies, such as mesh or multi-hop networks, to improve signal propagation over long distances and ensure consistent reliability \cite{wong2024multi}. Additionally, research should focus on dynamic spectrum management systems that adapt in real-time to environmental conditions and device density, ensuring stable long-range communication. The unpredictable environmental conditions of agriculture (e.g., dense vegetation, extreme weather, and uneven terrain) can disrupt signal transmission. Research into advanced error-correction algorithms, interference mitigation strategies, and resilient antenna designs can enhance the robustness of LoRa networks, ensuring reliable data transmission in such challenging environments. Furthermore, integrating AI into LoRa networks can improve real-time decision-making and dynamic optimization of communication protocols. AI models can predict network congestion, optimize resource allocation, and adapt to changing agricultural conditions, improving both the range and reliability of communication \cite{farhad2023ai,gia2019edge}. Research should also focus on integrating LoRa with other communication protocols (e.g., NB-IoT or Wi-Fi) to ensure seamless data transfer and reliability over long distances. Hybrid systems that balance low-power consumption with high data throughput can enable uninterrupted, reliable communication across extensive agricultural fields.

\subsection{LoRa-based ISAC for Agricultural IoT }
Future research on LoRa-based ISAC should prioritize the development of systems that enable simultaneous sensing and communication, optimizing real-time data collection across large farming areas \cite{ chang2023exploration}. These systems can enable seamless integration of environmental monitoring (e.g., soil moisture, crop health) with efficient data transmission, improving the precision and efficiency of agricultural operations. Moreover, integrating LoRa-based ISAC with autonomous farming machinery can enable real-time control and coordination of planting, irrigation, and harvesting operations. Additionally, future research should focus on creating resilient ISAC networks capable of predicting extreme weather events to enhance agricultural resilience by enabling farmers to take preventive actions.

Future research should also focus on incorporating blockchain technology into LoRa-based ISAC networks to ensure decentralized data integrity and security, preventing tampering and enhancing trust within farming ecosystems. LoRa-based ISAC should also focus on integrating edge computing to enable on-node data processing, reducing latency and bandwidth consumption, which is crucial for real-time tasks such as pest detection and precision irrigation. It is also necessary to explore enhancing bio-sensing LoRa nodes with ISAC capabilities, enabling real-time monitoring of plant health through biochemical indicators while simultaneously communicating the data for early interventions in cases of nutrient deficiencies or diseases \cite{gao2020framework}. Additionally, innovative approaches should apply ISAC to multi-crop farming systems, where distinct crop needs can be met by optimizing resource allocation and management for each type of crop. Future research should focus on developing LoRa-enabled micro-climate monitoring systems with ISAC capabilities for greenhouses, allowing precise control of temperature, humidity, and CO2 levels, thereby optimizing crop growth conditions in controlled environments \cite{codeluppi2020lorafarm}.

\section{Conclusion}\label{conclusions}
This survey paper has comprehensively explored the integration of LoRa technology in smart agriculture, emphasizing its pivotal role in advancing farming practices through IoT. By identifying a gap in existing literature regarding a communication-oriented review on LoRa technology for agriculture applications, this paper has highlighted the need for further research and development in this field.
The discussion on LoRa-based agriculture networks, including network architecture, PHY considerations, and tailored channel modeling, has shed light on the specifics of implementing LoRa technology in agricultural IoT systems. Additionally, examining relaying and routing mechanisms has addressed challenges in extending network coverage and optimizing data transmission in agricultural settings.
Practical implementation considerations, such as sensor deployment strategies and energy management techniques, have provided valuable insights into real-world deployment scenarios. Furthermore, the comparative analysis of LoRa with other wireless communication technologies commonly used in agriculture IoT has highlighted LoRa's advantages and limitations in this context.
Lastly, the outline of future research directions has identified areas for further exploration and innovation, aiming to leverage the potential of LoRa-based smart agriculture IoT systems fully. This survey paper is a valuable resource for researchers, technologists, and practitioners seeking to understand, implement, and advance smart agriculture initiatives using LoRa technology.
Future research in LoRa-based smart agriculture IoT can build on this paper's findings by delving into specific areas, such as advanced channel modeling techniques designed for heterogeneous farming environments and novel relay routing algorithms adapted for LoRa networks. In addition, exploring new sensor technologies and deployment methods can enhance the data collection and analysis of intelligent agricultural systems. This includes integrating emerging sensor techniques such as hyper-spectral imaging and drone-based sensing.
Future research efforts should focus on advancing our understanding of LoRa technology capabilities and limitations in smart agriculture, enabling more efficient, sustainable, and resilient agricultural practices for smart farming.


\bibliographystyle{IEEEtran}
\bibliography{mybibliography}

\end{document}